\documentclass[structabstract]{aa}  
\usepackage{graphicx}
\usepackage{amssymb}
\begin{document}

\title{The VLT/NaCo large program to probe the occurrence of exoplanets and brown dwarfs at wide
  orbits\thanks{Based on observations collected at the European
    Southern Observatory, Chile (ESO Large Program 184.C-0157 and Open
    Time 089.C-0137A and 090.C-0252A)}}
\subtitle{II- Survey description, results and performances}

\titlerunning{The VLT/NaCo large program for exoplanets at wide orbits II.}
\authorrunning{Chauvin et al.}

\author{G. Chauvin\inst{1}
        \and  A. Vigan\inst{2}
	\and M. Bonnefoy \inst{3}
	\and S. Desidera \inst{4}
	\and M. Bonavita \inst{4}
	\and D. Mesa \inst{4}
	\and A. Boccaletti \inst{5}
	\and E. Buenzli \inst{3}
	\and J. Carson\inst{6,3}
	\and P. Delorme \inst{1}
	\and J. Hagelberg \inst{7}
	\and G. Montagnier\inst{2}
	\and C. Mordasini \inst{3}
	\and S.P. Quanz \inst{8}
	\and D. Segransan \inst{7}
	\and C. Thalmann \inst{8}
	\and J.-L. Beuzit \inst{1}
	\and B. Biller \inst{3}
	\and E. Covino \inst{9}
	\and M. Feldt \inst{3}
	\and J. Girard \inst{10}
	\and R. Gratton \inst{4}
	\and T. Henning \inst{3}
	\and M. Kasper \inst{11}
	\and A.-M. Lagrange \inst{1}
	\and S. Messina \inst{12}
	\and M. Meyer \inst{8}
	\and D. Mouillet \inst{1}
	\and C. Moutou \inst{2}
	\and M. Reggianni  \inst{8}
	\and J.E. Schlieder \inst{3}
	\and A. Zurlo \inst{2}
}

\institute{
$^{1}$ UJF-Grenoble1/CNRS-INSU, Institut de Plan\'etologie et d'Astrophysique de Grenoble UMR 5274, Grenoble, F-38041, France\\
$^{2}$ Aix Marseille Universit\'e, CNRS, LAM (Laboratoire d'Astrophysique de Marseille) UMR 7326, 13388 Marseille, France\\
$^{3}$ Max Planck Institute for Astronomy, K\"onigstuhl 17, D-69117 Heidelberg, Germany\\
$^{4}$ INAF - Osservatorio Astronomico di Padova, Vicolo dell’ Osservatorio 5, 35122, Padova, Italy\\
$^{5}$ LESIA, Observatoire de Paris Meudon, 5 pl. J. Janssen, 92195 Meudon, France\\ 
$^{6}$ Department of Physics \& Astronomy, College of Charleston, 58 Coming Street, Charleston, SC 29424, USA\\ 
$^{7}$  Geneva Observatory, University of Geneva, Chemin des Mailettes 51, 1290
Versoix, Switzerland\\
$^{8}$  Institute for Astronomy, ETH Zurich, Wolfgang-Pauli-Strasse 27, 8093 Zurich, Switzerland\\
$^{9}$  INAF Osservatorio Astronomico di Capodimonte Via Moiarello 16 80131 Napoli Italy\\
$^{10}$  European Southern Observatory, Casilla 19001, Santiago 19, Chile\\
$^{11}$  European Southern Observatory, Karl Schwarzschild St, 2, D-85748 Garching, Germany \\
$^{12}$  INAF - Catania Astrophysical Observatory, via S. So a 78 I-95123 Catania, Italy\\
}
  \abstract
     {Young, nearby stars are ideal targets for direct imaging
       searches for giant planets and brown dwarf companions. After the
       first imaged planets discoveries, vast
       efforts have been devoted to the statistical analysis of the
       occurence and orbital distributions of giant planets and brown
       dwarf companions at wide ($\ge5-6$~AU) orbits.}
   {In anticipation of the VLT/SPHERE planet imager guaranteed time
     programs, we have conducted a preparatory survey of 86 stars
     between 2009 and 2013 in order to identify new faint comoving
     companions to ultimately carry out a comprehensive analysis of
     the occurence of giant planets and brown dwarf companions at wide
     ($10-2000$~AU) orbits around young, solar-type stars.}
     {We used NaCo at VLT to explore the occurrence rate of giant
       planets and brown dwarfs between typically 0.1 and
       $8~\!''$. Diffraction-limited observations in $H$-band combined
       with angular differential imaging enabled us to reach primary
       star-companion brightness ratios as small as $10^{-6}$ at
       $1.5~\!''$. Repeated observations at several epochs enabled us
       to discriminate comoving companions from background objects.}
    {12 systems were resolved as new binaries, including the discovery
      of a new white dwarf companion to the star HD\,8049. Around 34
      stars, at least one companion candidate was detected in the
      observed field of view. More than 400 faint sources were
      detected, 90\% of them in 4 crowded fields. With the exception
      of HD\,8049\,B, we did not identify any new comoving
      companions. The survey also led to spatially resolved images of
      the thin debris disk around HD\,61005 that have been published
      earlier. Finally, considering the survey detection limits, we
      derive a preliminary upper limit on the frequency of giant
      planets for semi-major axes of [10,2000]~AU: typically less than
      $15\%$ between 100 and 500~AU, and less than $10\%$ between 50
      and 500~AU for exoplanets more massive than 5~M$_{\rm{Jup}}$ and
      10~M$_{\rm{Jup}}$ respectively, considering a uniform input
      distribution and with a confidence level of 95\%.}
     {The results from this survey are in agreement with earlier
       programs emphasizing that massive, gas giant companions on wide
       orbits around solar-type stars are rare. They will be part of a
       broader analysis of a total of $\sim210$ young, solar-type
       stars to bring further statistical constraints for theoretical
       models of planetary formation and evolution.}

   \keywords{Instrumentation: adaptive optics, high angular resolution -- Methods: observational, statistical -- Techniques: image processing -- Surveys -- Stars: close binaries, planetary systems, brown dwarfs -- Infrared: stars, planetary systems}

   \maketitle
%

\section{Introduction}


Our understanding of the origin and evolution of extrasolar planets
(EPs) has drastically transformed in the last decade.  Current
theories favor the formation of planets within a protoplanetary disk
by accretion of solids, building up a 10 to 15~M$_{\oplus}$ core
followed by rapid agglomeration of gas (Pollack et al. 1996; Alibert
et al. 2004), or by gravitational instability of the gas (Boss 1997;
Stamatellos \& Withworth 2008; Vorobyov 2013). Whereas physical conditions
and timescales favor core accretion in the inner disk ($\le10$~AU),
gravitational instability could be the main mechanism to form massive
gaseous giants at wider separations ($\ge10$~AU) in the earliest phase
of the disk's lifetime (Boley 2009). The planets could either migrate
inward, toward or outward, from the star by disk-planet interactions
(Kley \& Nelson 2012 and reference therein) or during planet-planet
interactions (Naoz et al. 2011; Dawson \& Murray-Clay 2013), which
will alter the original semi-major axis distribution. A wide range of
potential planet masses, sizes, locations and compositions results
from this flurry of formation and evolution possibilities.  A major
goal for exoplanetary science of the next decade is a better
understanding of these mechanisms. In this context, the role of
observations is crucial to provide constraints that will help to model
the diversity of exoplanetary properties. The main observables are the
occurrence of EPs, the physical and orbital characteristics
(composition, mass, radius, luminosity, distribution of mass, period
and eccentricity), but also the properties of the planetary hosts
(mass, age, metallicity, lithium abundance or multiplicity).
 
Brown dwarfs (BDs) were originally proposed as a distinguishable class
of astrophysical objects, with intermediate masses between stars and
planets. Recent large infrared surveys and high contrast observations
have unambiguously revealed the existence of planetary mass objects,
isolated in the field (Zapatero-Osorio et al. 2000; Liu et al. 2013; Joergens
et al. 2013) or wide companions to stars (Chauvin et al. 2005a). Their
existence confirms that the formation mechanisms proposed to form
stars (gravo-turbulent fragmentation, disk fragmentation,
accretion-ejection or photo-erosion; see Whitworth et al. 2007, Luhman
2012 for reviews) can actually form objects down to the planetary mass
regime. The details of contraction and subsequent evolution of the
cores remain critical and are still under considerable
debate. Episodic accretion processes can affect their physical
properties (Baraffe et al. 2009). It is now undeniable that the
stellar and planetary formation mechanisms overlap in the substellar
regime. They can both lead to the formation of planetary mass objects,
including companions to stars and BDs. Fossil traces of the formation
processes should be revealed by different physical features (presence
of core, composition of the atmosphere, system
architecture...). Distinct statistical properties such as the
occurrence, the mass, separation and eccentricity distributions,
should help to identify the dominant mechanism to form substellar
companions.

The main statistical constraints on exoplanets originally came from
the radial velocity (RV) technique. More than 800 EPs have been now
confirmed, featuring a broad range of physical (mass) and orbital (P,
$e$) characteristics around different stellar hosts (Howard et
al. 2010; Mayor et al. 2011; Wright et al. 2012; Bonfils et
al. 2013). The strong bimodal aspect of the secondary-mass
distribution to solar-type primaries has generally been considered the
most obvious evidence of different formation mechanisms for stellar
and planetary systems. The period distribution of giant exoplanets is
basically made of two main features: a peak around 3 days plus an
increasing frequency as a function of period (Udry \& Santos
2007). The observed pile up of planets with periods around 3 days is
believed to be the result of migration and final stopping
mechanism. The rise of the number of planets with increasing distance
from the parent star reaches up to a separation corresponding to the
duration limit of most of the longest surveys (5-6~AU). This
extrapolation hints that a large population of yet undetected
Jupiter-mass planets may exist beyond 5~AU, suggesting an ideal niche
for the direct-imaging surveys. More recently, a plethora of
transiting planetary candidates have been revealed by \textit{Kepler}
(more than 2300 candidates known today, Batalha et al. 2013), probably
corroborating how abundant telluric planets are and in agreement with
Doppler surveys in terms of occurrence at less than 0.25~AU (Howard et
al. 2012).

\begin{table*}[t]
\begin{centering}
\label{tab:deepsurveys}
\caption{Deep imaging surveys of young ($<100$~Myr) and
  intermediate-old to old ($0.1-5$~Gyr), nearby ($<100$~pc) stars
  dedicated to the search for planetary mass companions. We have indicated the telescope and the instrument, the imaging mode
  (Cor-I: coronagraphic imaging; Sat-I; saturated imaging; I: imaging;
  SDI: simultaneous differential imaging; ADI: angular differential
  imaging; ASDI: angular and spectral differential imaging), the
  filters, the field of view (FoV), the number of stars observed (\#),
  their spectral types (SpT) and ages (Age). }
\begin{tabular}{lllllllllll}     
\hline\noalign{\smallskip}
Reference               & Telescope        & Instr.       &  Mode          & Filter       & FoV          & \#       & SpT      & Age       \\ 
                        &                  &              &                &              & (  $''\times''$)     &          &          & (Myr)     \\
\noalign{\smallskip}\hline\noalign{\smallskip}
Chauvin et al. 2003     & ESO3.6m          & ADONIS       & Cor-I          & $H, K$        & $13\times13$ & 29      & GKM      & $\lesssim50$  \\ 
Neuh\"auser et al. 2003  & NTT              & Sharp        & Sat-I          & $K$          & $11\times11$ & 23       & AFGKM     & $\lesssim50$  \\ 
                        & NTT              & Sofi         & Sat-I          & $H$          & $13\times13$ & 10       & AFGKM     & $\lesssim50$  \\ 
Lowrance et al. 2005    & HST              & NICMOS       & Cor-I          & $H$          & $19\times19$ & 45       & AFGKM    & $10-600$  \\
Masciadri et al. 2005   & VLT              & NaCo         & Sat-I          & $H, K$        & $14\times14$ & 28       & KM       & $\lesssim200$  \\ 
Biller et al. 2007      & VLT              & NaCo         & SDI            & $H$          & $5\times5$   & 45       & GKM      & $\lesssim300$  \\
                        & MMT              &              & SDI            & $H$          & $5\times5$   & -         & -         & -          \\ 
Kasper et al. 2007      & VLT              & NaCo         & Sat-I          & $L'$         & $28\times28$ & 22       & GKM      & $\lesssim50$  \\ 
Lafreni\`ere et al. 2007& Gemini-N         & NIRI         & ADI            & $H$          & $22\times22$ & 85   &          & 10-5000   \\
Apai et al. 2008$^a$        & VLT              & NaCo         & SDI            & $H$          & $3\times3$   & 8    &    FG & 12-500 \\
Chauvin et al. 2010     & VLT              & NaCo         & Cor-I          & $H, K$        & $28\times28$ & 88       & BAFGKM    & $\lesssim100$          \\  
Heinze et al. 2010ab    & MMT              & Clio         & ADI            & $L', M$       & $15.5\times12.4$ & 54   & FGK      & 100-5000  \\                       
Janson et al. 2011      & Gemini-N         & NIRI         & ADI            & $H, K$        & $22\times22$ & 15       & BA       & 20-700    \\  
Vigan et al. 2012       & Gemini-N         & NIRI         & ADI            & $H, K$        & $22\times22$ & 42       & AF       & 10-400    \\  
                        & VLT              & NaCo         & ADI            & $H, K$        & $14\times14$ & -         &  -        & -          \\
Delorme et al. 2012     & VLT              & NaCo         & ADI            & $L'$         & $28\times28$ & 16       & M        & $\lesssim200$  \\  
Rameau et al. 2013c      & VLT              & NaCo         & ADI            & $L'$         & $28\times28$ & 59       & AF       & $\lesssim200$  \\
Yamamoto et al. 2013       & Subaru           & HiCIAO       & ADI            & $H, K$          & $20\times20$ & 20       & FG       & $125\pm8$  \\
Biller et al. 2013      & Gemini-S         & NICI         & Cor-ASDI       & $H$          & $18\times18$ & 80       & BAFGKM     &  $\lesssim200$    \\ 
Brandt et al. 2013$^b$      & Subaru           & HiCIAO       & ADI            & $H$          & $20\times20$ & 63       & AFGKM    & $\lesssim500$    \\ 
Nielsen et al. 2013     & Gemini-S         & NICI         & Cor-ASDI       & $H$          & $18\times18$ & 70       & BA       & 50-500    \\ 
Wahhaj et al. 2013$^a$      & Gemini-S         & NICI         & Cor-ASDI       & $H$          & $18\times18$ & 57       & AFGKM    & $\sim100$ \\ 
Janson et al. 2013$^a$      & Subaru           & HiCIAO       & ADI            & $H$          & $20\times20$ & 50       & AFGKM    & $\lesssim1000$ \\  
 
\noalign{\smallskip}\hline                  
\end{tabular}
\begin{list}{}{}
\item[\scriptsize{- ($^a$):}] \scriptsize{surveys dedicated to planets around debris disk stars.}
\item[\scriptsize{- ($^b$):}] \scriptsize{paper submitted.}
\end{list}
\end{centering}
\end{table*}

Despite the success of the RV and transit techniques, the time spans
explored limit the studies to the close ($\le 5-6$~AU) EPs. Within the
coming years, direct imaging represents the only viable technique for
probing the existence of EPs and BD companions at large ($\ge 5-6$~AU)
separations. This technique is also unique for the characterization of
planetary atmospheres that are not strongly irradiated by the
planetary host (Janson et al. 2010; Bowler et al. 2010; Barman et
al. 2011a, 2011b; Bonnefoy et al. 2010, 2013a, 2013b, 2013c; Konopacky
et al. 2013).  Young ($\le500$~Myr), nearby stars are very favourable
targets for the direct detection of the lowest mass companions.  Since
the discovery of the TW Hydrae association (TWA) by Kastner et
al. (1997) and Hoff et al. (1998), more than 300 young, nearby stars
were identified. They are gathered in several groups (TWA, $\beta$
Pictoris, Tucana-Horologium, $\eta$ Cha, AB Dor, Columba, Carinae),
sharing common kinematics, photometric and spectroscopic properties
(see Zuckerman \& Song 2004; Torres et al. 2008). With typical
contrast of $10-15$ magnitudes for separations beyond $1.0-2.0~\!''$
($50-100$ AU for a star at 50 pc), planetary mass companions down to
$1-2$ Jupiter masses are potentially detectable by current very deep
imaging surveys. The first planetary mass companions were detected at
large distances ($\ge100$~AU) and/or with small mass ratio with their
primaries, indicating a probable star-like or gravitational disk
instability formation mechanism (Chauvin et al. 2005b; Lafreni\`ere et
al. 2008).

The breakthrough discoveries of closer and/or lighter planetary mass
companions like Fomalhaut~b ($<1$~M$_{\rm{Jup}}$ at 177~AU; Kalas et
al. 2008, 2013), HR\,8799\,bcde (10, 10, 10 and 7~M$_{\rm{Jup}}$ at
resp. 14, 24, 38 and 68~AU; Marois et al. 2008, 2010), $\beta$
Pictoris b (8~M$_{\rm{Jup}}$ at 8~AU; Lagrange et al. 2009) or more
recently $\kappa$~And\,b ($14^{+25}_{-2}$~M$_{\rm{Jup}}$ at 55~AU;
Carson et al. 2013; Bonnefoy et al. 2013c), HD\,95086\,b
($4-5$~M$_{\rm{Jup}}$ at 56~AU; Rameau et al. 2013a, 2013b) and
GJ\,504\,b ($4^{+4.5}_{-1}$~M$_{\rm{Jup}}$ at 43.5~AU; Kuzuhara et
al. 2013) indicate that we are just initiating the characterization of
the outer part of planetary systems between typically $5-100$~AU. Vast
efforts are now devoted to systematic searches of EPs in direct
imaging with an increasing number of large scale surveys (see Table~1;
nine new surveys published between 2012 and 2013). The number of
targets surveyed and the detection performances will increase with the
new generation of planet finders LMIRCam at LBT (Skrustkie et
al. 2010), MagAO (Close et al. 2012), ScExAO at Subaru (Guyon et
al. 2010), SPHERE at VLT (Beuzit et al. 2008), GPI at Gemini
(Macintosh et al. 2008) with the goal to provide better statistics on
larger samples and a greater number of giants planets to be
characterized. It should enable to test alternative mechanisms to the
standard planetary formation theories of core accretion and
gravitation instability such as pebble accretion (Lambrechts \&
Johansen 2012; Morbidelli \& Nesvorny 2012) or tidal downsizing (Boley
et al. 2010; Nayakshin 2010; Forgan \& Rice 2013) that are currently
proposed to explain the existence of a population of giant planets at
wide orbits. In the context of the VLT/SPHERE scientific preparation,
we have conducted a large observing program (ESO: 184.C-0157) of 86
stars with NaCo (hereafter the NaCo-LP). Combined with stars already
observed in direct imaging, it represents a total of more than
$\sim210$ stars to study the occurrence rate of giant planets and
brown dwarf companions at wide ($10-2000$~AU) orbits. This complete
analysis is detailed in series of four papers: a description of the
complete sample (Desidera et al. 2013, submitted), the NaCo-LP survey
(this paper) and the statistical analysis of the giant planet
population (Vigan et al. 2014, in prep) and of the brown dwarf
companion population (Reggianni et al. 2014, in prep). We therefore
report here the results of the NaCo-LP carried out between 2009 and
2013. In Section~2, we describe the target sample selection. In
Section~3, we detail the observing setup. In Section~4, the data
reduction strategy and analysis are reported with the results in
Section~5. Finally, a preliminary statistical analysis of the observed
sample is presented in Section~6 and our main conclusions in
Section~7.

\begin{figure*}[t]
\hspace{0.2cm}
\includegraphics[width=5.6cm]{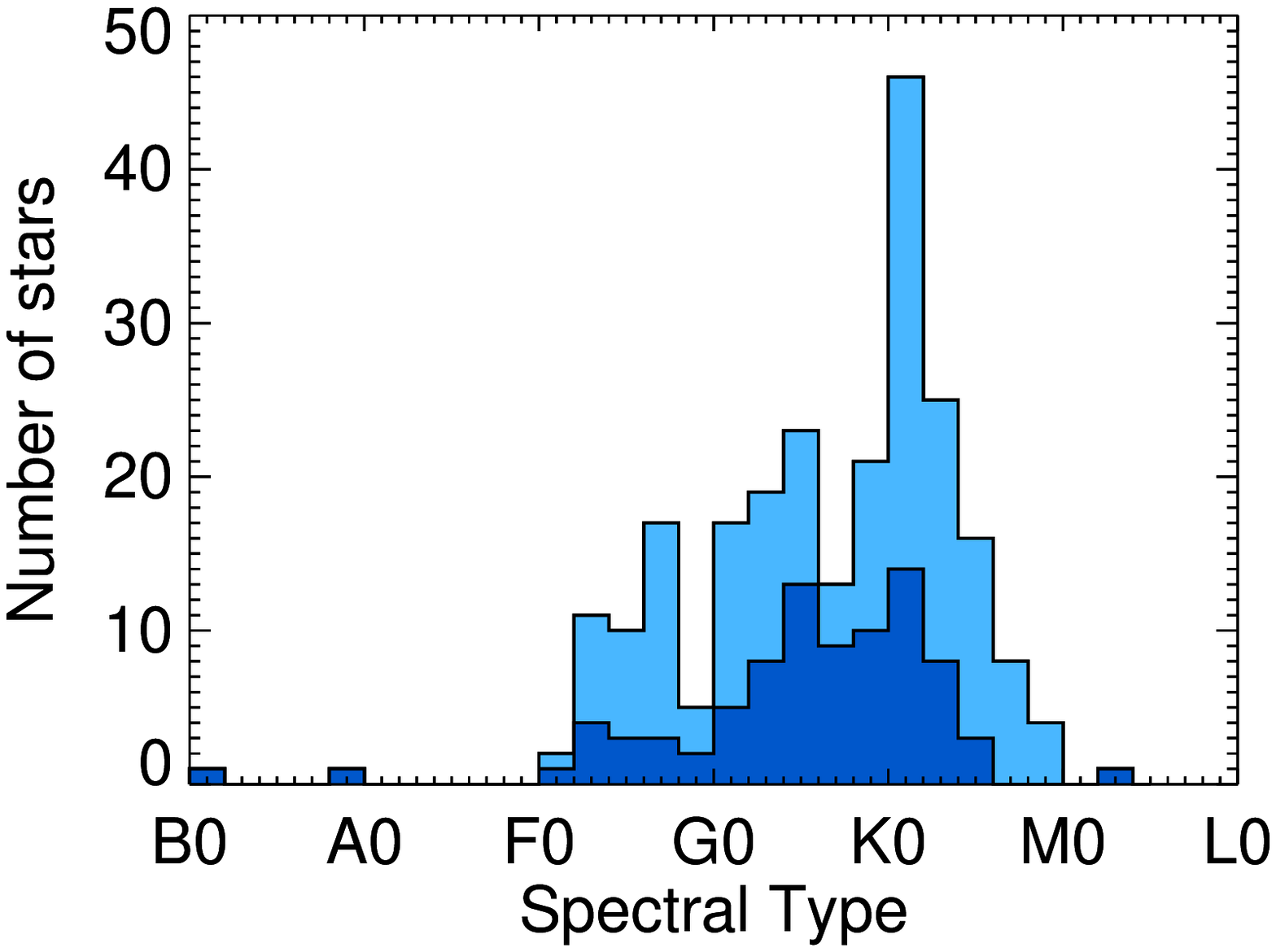}
\includegraphics[width=5.6cm]{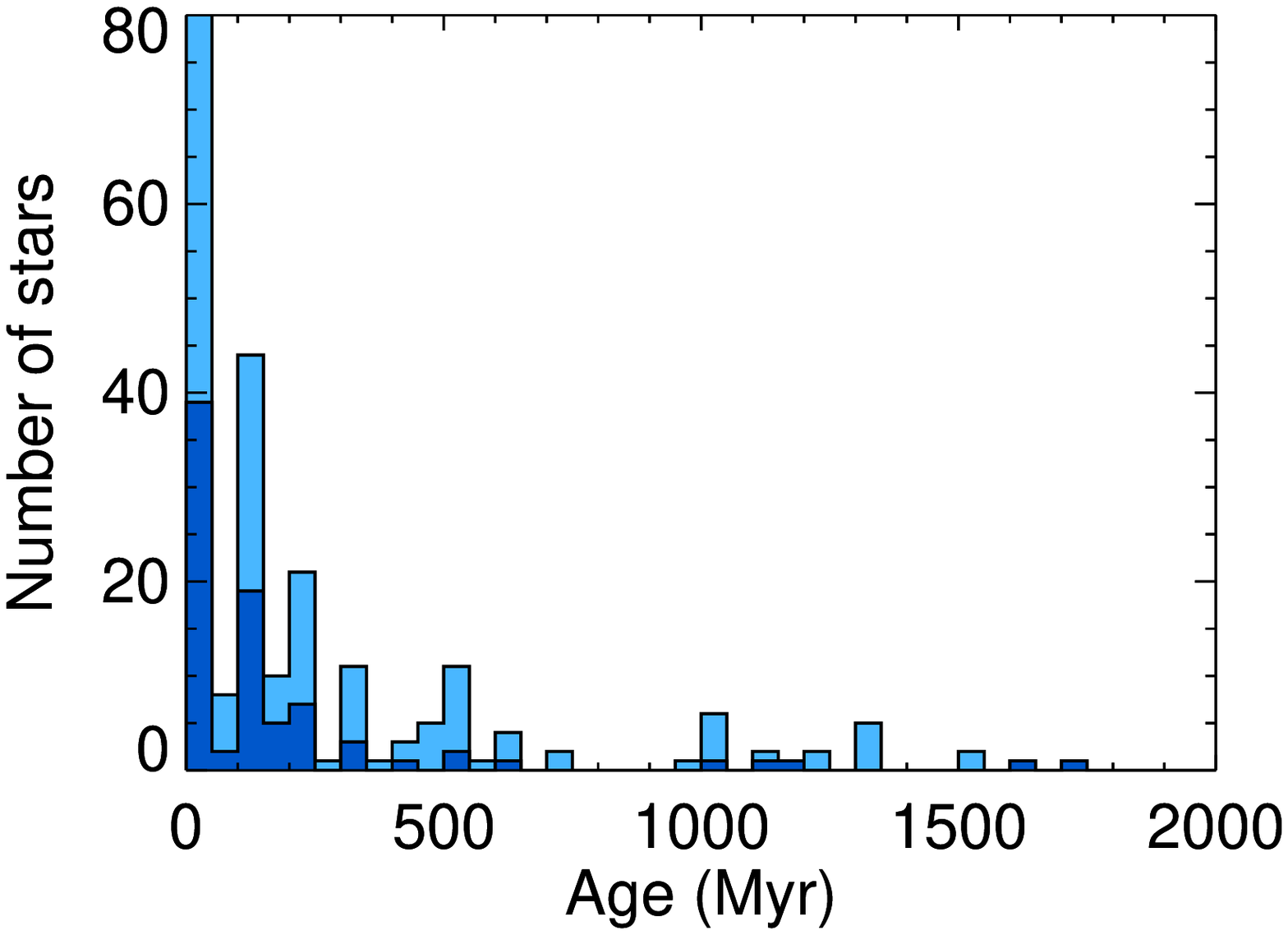}
\includegraphics[width=5.6cm]{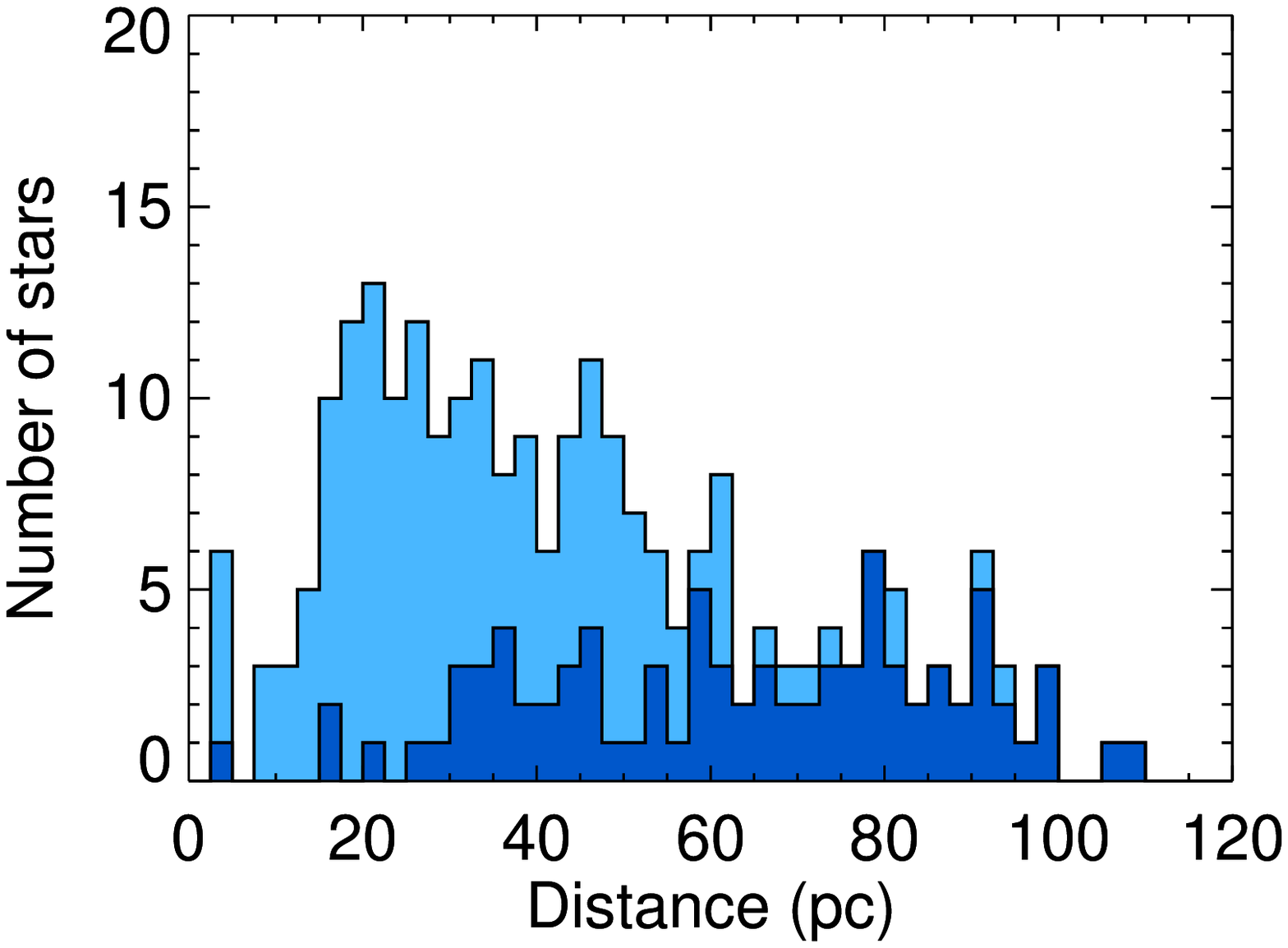}\\
\includegraphics[width=5.6cm]{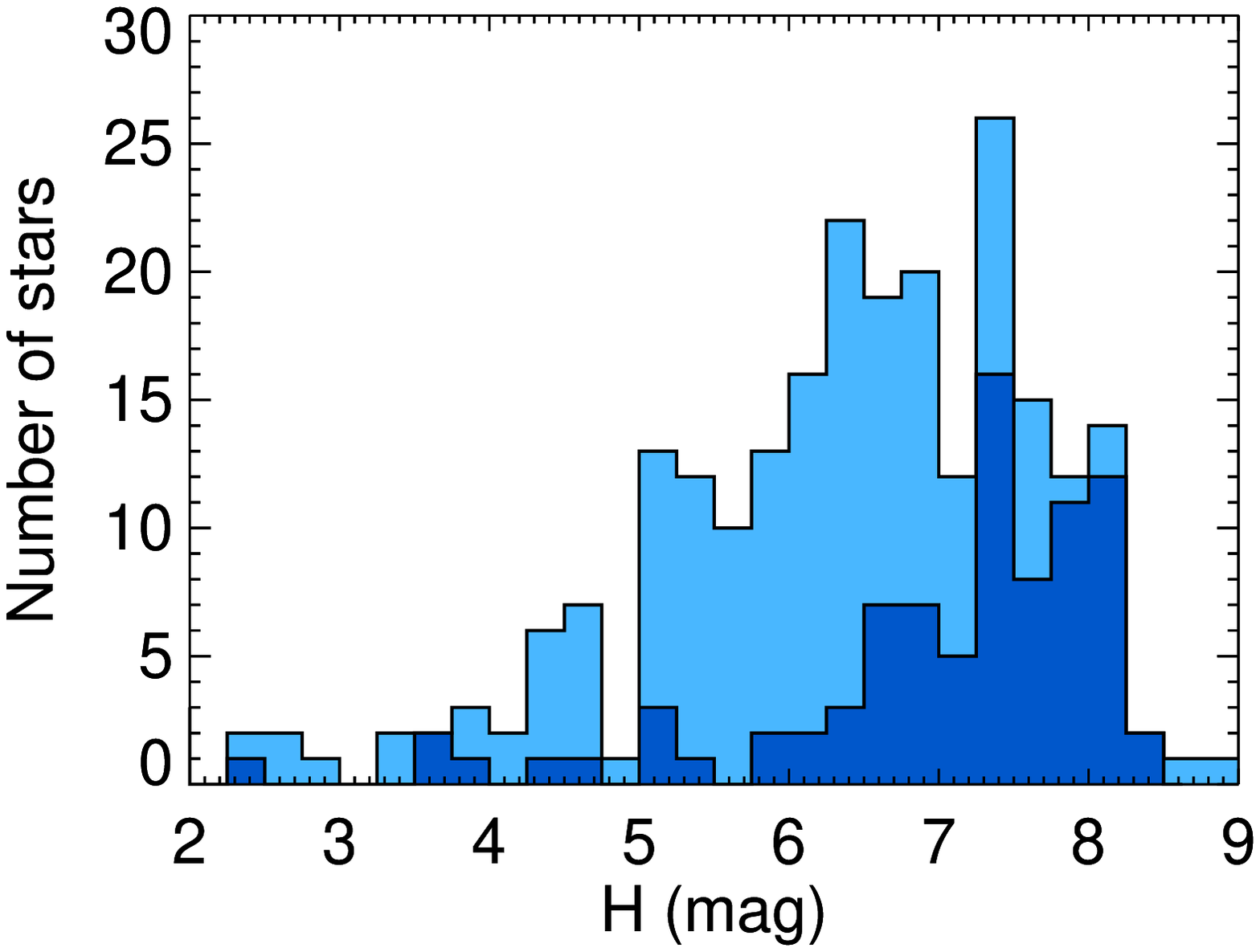}
\includegraphics[width=5.6cm]{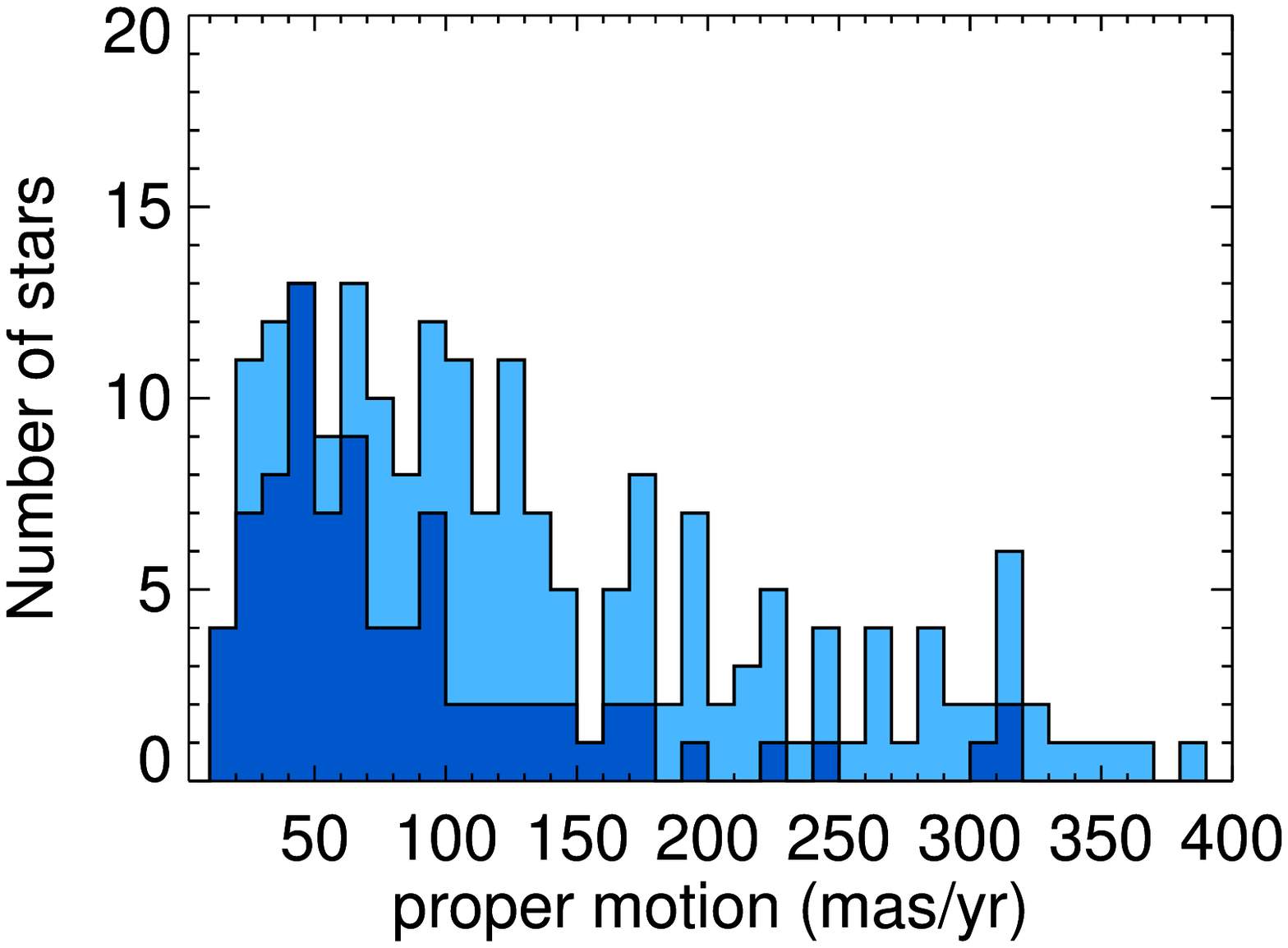}
\includegraphics[width=5.6cm]{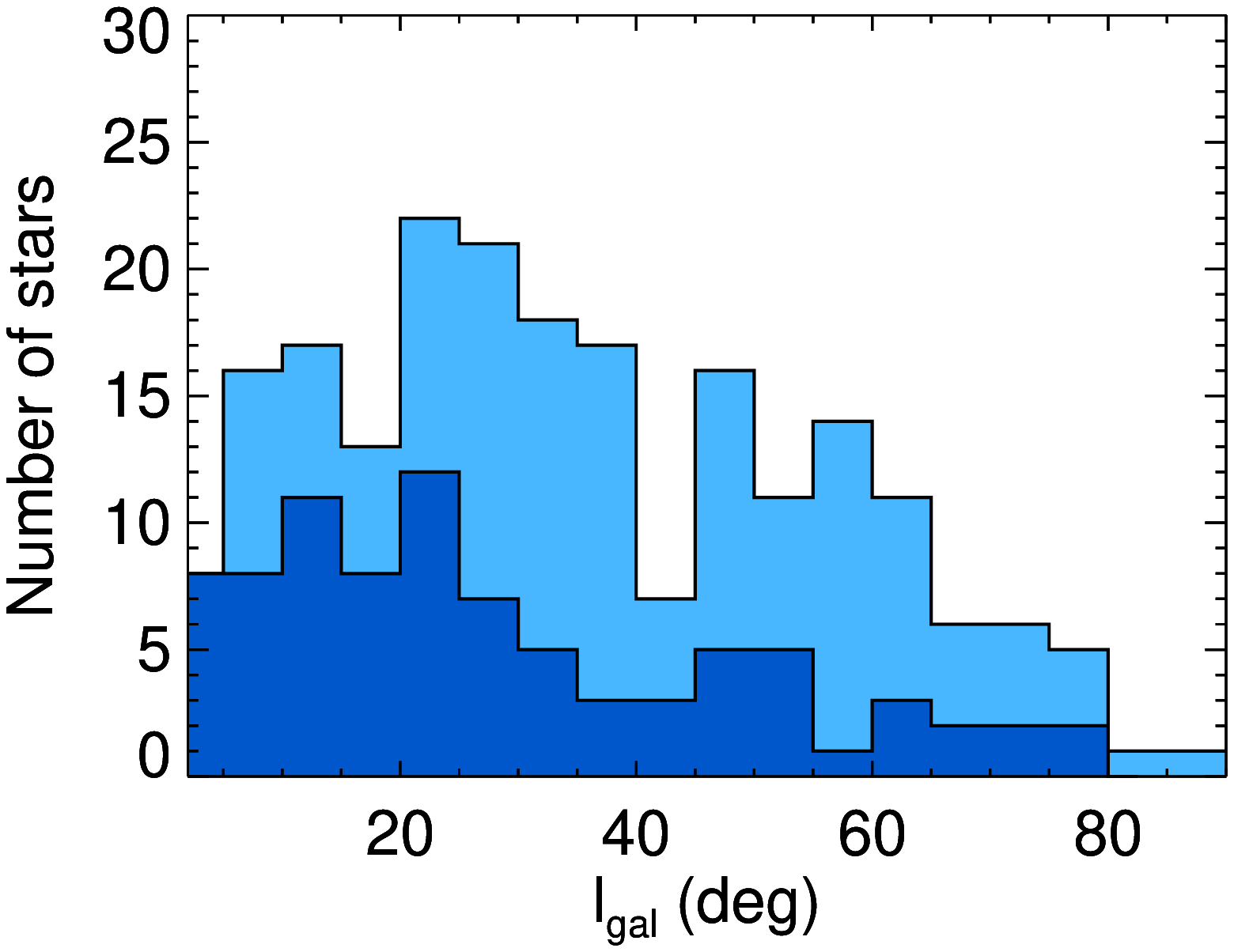}
\begin{centering}

\caption{Histrograms summarizing the main properties of NaCo-LP
  observed sample (\textit{dark blue}) and of the final NaCo-LP
  statistical sample of $\sim$210 stars (\textit{light blue}) used by
  Vigan et al. (2014, in prep) and Reggianni et al. (2014, in prep):
  spectral type, age, distance, $H$-band magnitude, proper motion
  amplitude and galactic latitude.}

\label{fig:sample_prop}
\end{centering}
\end{figure*}

\begin{table*}[!t]

\caption{NaCo-LP target sample and properties. In addition to the
  target name, H-band magnitude, spectral type, distance and age, we
  have reported the multiplicity status with a flag (Bin for visual
  binaries with the indication that they are new or known; SB for
  spectroscopic binaries; RV var for radial velocity variable), the
  observing mode (nonsat for non-saturated or sat for saturated, ADI
  for angular differential imaging and the filter) and the presence of
  companion candidates (ccs).}

\begin{center}
\small
\label{tab:sample1}      
\begin{tabular}{lllllllll}     
\noalign{\smallskip}
\noalign{\smallskip}\hline
Name-1    &    Name-2    &    H     &     SpT     &      d      &     Age    &  Binarity   &    Mode     &    Comments     \\
          &              & (mag)    &             & (pc)        & (Myr)      &             &             &                 \\
\noalign{\smallskip}\hline  \noalign{\smallskip}
 TYC\,5839-0596-1 &        BD-16-20 &     6.6&    K0IVe &    43.5&   150.       & SB2 & sat, ADI, H &  \\
 TYC\,0603-0461-1 &        BD+07-85 &     7.4&     K4Ve &    58.2&   100.       & Bin (new) & sat, ADI, H &  \\
          HIP3924 &          HD4944 &     6.7&      F7V &    53.2&  500.       & SB2 & sat, ADI, H &  \\
          HIP6177 &          HD8049 &     6.7&      K2V &    33.6&  3000.       & & sat, ADI, H &  cc \\
          HIP8038 &        HD10611B &     7.2&     K5Ve &    29.3&   150.       & Bin (new) & nonsat, ADI, H &  \\
         HIP10602 &         HD14228 &     4.0&      B0V &    47.1&    30.       & & sat, ADI, H &  \\
         HIP11360 &         HD15115 &     5.9&       F2 &    45.2&    30.       & & sat, ADI, H &  \\
 TYC\,8484-1507-1 &       CD-53-535 &     6.6&      G8V &    60.5&   100.       & Bin (known) & nonsat, ADI, H &  \\
         HIP12394 &         HD16978 &     4.4&    B9III &    46.6&    30.       & & sat, ADI, H &  \\
         HIP13008 &         HD17438 &     5.5&      F2V &    39.6&   1000.       & & sat, ADI, H & cc \\
         HIP14684 &          IS-Eri &     6.8&       G0 &    37.4&   100.       & & sat, ADI, H &  cc \\
 TYC\,8060-1673-1 &      CD-46-1064 &     7.2&      K3V &    40.4&    30.       & & sat, ADI, H &  \\
         HIP19775 &         HD26980 &     7.7&      G3V &    80.5&    30.       & & sat, ADI, H &  \\
         HIP23316 &         HD32372 &     7.9&      G5V &    76.3&    30.       & & sat, ADI, H &  \\
          HD32981 &      BD-16-1042 &     7.8&      F8V &    86.7&   100.       & & sat, ADI, H &  \\
      BD-09-1108 &             xxxx &     8.2&       G5 &    93.6&    30.       & & sat, ADI, H &  \\
       HIP25434 &          HD274197 &     7.9&       G0 &    79.1&    20.       & & sat, ADI, H &  ccs \\
 TYC\,9162-0698-1 &        HD269620 &     8.2&      G6V &    77.7&    30.       & & sat, ADI, H &  ccs \\
  TYC\,5346-132-1 &      BD-08-1195 &     8.1&       G7 &    81.2&    30.       & & sat, ADI, H &  ccs \\
       HIP30261 &           HD44748 &     7.6&      G6V &    61.8&   100.       & & sat, ADI, H &  \\
 TYC\,7617-0549-1 &      CD-40-2458 &     8.2&      K0V &    77.8&    30.       & & sat, ADI, H &  cc \\
 TYC\,9181-0466-1 &         HD47875 &     7.4&      G4V &    77.7&    30.       & Bin (new) & nonsat, ADI, H &  \\
       HIP32235 &           HD49855 &     7.4&      G6V &    58.2&    30.       & & sat, ADI, H &  cc \\
       HIP35564 &           HD57852 &     5.1&       F2 &    31.7&   200.       & RV var & sat, ADI, H &  ccs \\
 TYC\,8128-1946-1 &      CD-48-2972 &     8.1&      G8V &    89.7&    45.       & & sat, ADI, H &  ccs \\
       HIP36414 &           HD59704 &     6.5&      F7V &    52.5&   200.       & SB, RV var & sat, ADI, H &  ccs \\
       HIP36948 &           HD61005 &     6.6&      G5V &    35.3&    45.       & & sat, ADI, H &  ccs \\
       HIP37563 &           HD62850 &     5.9&      G3V &    32.8&   200.       & & sat, ADI, H &  \\
       HIP37923 &           HD63608 &     6.5&      K0V &    36.8&   200.       & & sat, ADI, H &  ccs \\
 TYC\,8927-3620-1 &         HD77307 &     7.7&     G8IV &    81.8&   20.       & Bin (new) & sat, ADI, H &  \\
       HIP46634 &       BD+11-2052B &     6.8&       G5 &    42.3&   300.       & & sat, ADI, H &  \\
       HIP47646 &           HD84199 &     6.9&      F5V &    73.6&   1150.       & & sat, ADI, H &  \\
          TWA-21 &         HD298936 &     7.3&     K3Ve &    54.8&    17.       & & sat, ADI, H &  ccs \\
 TYC\,7188-0575-1 &      CD-31-8201 &     7.4&     K0Ve &    43.2&   150.       & SB2 & sat, ADI, H &  ccs \\
 TYC\,6069-1214-1 &      BD-19-3018 &     8.0&      K0V &    67.8&    70.       & & sat, ADI, H &   \\
 TYC\,7722-0207-1 &        HD296790 &     7.8&      K0V &    65.8&   100.       & & sat, ADI, H &  ccs \\
 TYC\,7743-1091-1 &         HD99409 &     5.2&    G6III &   200.0&  1700.       & & sat, ADI, H &  \\
       HIP58240 &          HD103742 &     6.2&      G3V &    31.8&   200.       & & sat, ADI, H &  cc \\
 TYC\,9231-1566-1 &        HD105923 &     7.3&      G8V &    96.0&    10.       & Bin (new) & nonsat, ADI, H &  \\
 TYC\,8979-1683-1 &       CD-62-657 &     7.5&      G7V &    75.6&    17.       & & sat, ADI, H &  ccs \\
 TYC\,8989-0583-1 &        HD112245 &     7.4&     K0Ve &    65.4&    17.       & Bin (new) & sat, ADI, H &  \\
 TYC\,9245-0617-1 &      CD-69-1055 &     7.7&     K0Ve &    93.0&    10.       & & sat, ADI, H &  ccs \\
       HIP63862 &          HD113553 &     6.8&      G5V &    49.0&   150.       & & sat, ADI, H &  ccs \\
 TYC\,7796-2110-1 &      CD-41-7947 &     8.3&    K2IVe &    92.1&    17.       & & sat, ADI, H &  ccs \\
 TYC\,9010-1272-1 &        HD124831 &     7.8&      G3V &    86.5&    30.       & Bin (new) & nonsat, ADI, H &  \\
       HIP70351 &          HD125485 &     7.6&      G7V &    91.7&   110.       & & sat, ADI, H &  ccs \\
       HIP71908 &           GJ-560A &     2.5&      F1V &    16.6&  1110.       & & sat, ADI, H &  cc\\
       HIP71933 &          HD129181 &     7.2&      F8V &    83.9&    16.       & & sat, ADI, H &  ccs \\
       HIP72399 &         HD130260A &     7.5&     K3Ve &    46.1&   500.       & SB1, RV var & sat, ADI, H &  \\
 TYC\,7835-2569-1 &        HD137059 &     7.1&      G3V &    70.2&   120.       & SB2 + Bin (known) & sat, ADI, H &  \\
      HIP76829     &       HD139664 &     3.7&     F5IV &    17.4&   200.       & & sat, ADI, H &  ccs \\
 TYC\,6781-0415-1 &     CD-24-12231 &     7.4&    G9IVe &   106.0&    11.       & & sat, ADI, H &  \\
 TYC\,6786-0811-1 &     CD-27-10549 &     7.5&     K0IV &    78.6&    60.       & Bin (known) & nonsat, ADI, H &  \\
       HIP78747    &       HD143928 &     5.1&      F3V &    37.9&  1600.       & & sat, ADI, H &  ccs \\
 TYC\,6209-0769-1 &      BD-19-4341 &     7.4&     K0IV &    43.9&   120.       & & sat, ADI, H &  cc \\
       HIP79958    &       HD146464 &     6.7&     K3Ve &    27.2&   130.       & & sat, ADI, H &  ccs \\
       HIP80290    &       HD147491 &     8.0&     G2IV &    83.3&    30.       & Bin (new) & sat, ADI, H &  ccs \\
       HIP80758    &       HD148440 &     8.0&     G9Ve &    98.2&    20.       & & sat, ADI, H &  ccs \\
 \noalign{\smallskip}\hline  \noalign{\smallskip}
\end{tabular}
\end{center}
\end{table*}

\begin{table*}[t]
\caption{NaCo-LP target sample and properties (Table~2-cont).}             
\label{tab:sample2}      
\begin{center}
\small
\begin{tabular}{lllllllll}     
\noalign{\smallskip}
\noalign{\smallskip}\hline
Name-1    &    Name-2    &    H     &     SpT     &      d      &     Age    &  Binarity   &    Mode     &    Comments     \\
          &              & (mag)    &             & (pc)        & (Myr)      &             &             &                 \\
\noalign{\smallskip}\hline  \noalign{\smallskip}
TYC\,6818-1336-1   &       HD153439 &     7.8&     G0IV &    89.5&    30.       & & sat, ADI, H & ccs \\
 TYC\,6815-0084-1  &    CD-25-11942 &     7.7&     K0IV &    92.0&    11.       & & sat, ADI, H &  ccs \\
TYC\,6815-0874-1   &    CD-25-11922 &    10.1 &    G2IV &    109.0&    20         & SB2? & sat, ADI, H & ccs \\
 TYC\,7362-0724-1  &       HD156097 &     7.8&      G5V &    90.0&    20.       & & sat, ADI, H &  ccs \\
 TYC\,8728-2262-1  &     CD-54-7336 &     7.5&      K1V &    70.4&    12.       & & sat, ADI, H &  ccs \\
       HIP86672    &       HD160682 &     7.4&      G5V &    78.0&    30.       & & sat, ADI, H &  ccs \\
       HIP89829    &       HD168210 &     7.2&      G5V &    72.6&    16.       & & sat, ADI, H &  ccs \\
       HIP93375    &       HD176367 &     7.3&      G1V &    58.8&    100.       & & sat, ADI, H &  ccs \\
       HIP94235    &       HD178085 &     7.0&      G1V &    61.3&    100.       & Bin (new) & nonsat, ADI, H &  \\
  TYC\,6893-1391-1 &     CD-25-14224&     7.8&      K2V &    55.1&   160.       & & sat, ADI, H &  ccs\\
 TYC\,5206-0915-1  &      BD-07-5533&     8.2&      K1IV&    76.4&   300.       & & sat, ADI, H &  \\ 
 TYC\,5736-0649-1  &      BD-14-5534&     8.0&      G6V &    86.4&    30.       & & sat, ADI, H &  ccs\\
       HD189285    &      BD-04-4987&     8.0&       G5 &    77.8&   100.       & & sat, ADI, H &  cc \\
       HIP98470    &       HD189245 &     4.6&      F7V &    21.2&   100.       & & sat, ADI, H &  \\
  TYC\,5164-567-1  &      BD-03-4778&     8.0&          &    63.3&   100.       & & sat, ADI, H &  ccs \\
       HIP99273    &       HD191089 &     6.1&      F5V &    52.2&    16.       & & sat, ADI, H &  \\
       HD199058    &      BD+08-4561&     7.0&       G5 &    66.2&   100.       & Bin (new) & nonsat, ADI, H  &  \\
      HIP105384    &       HD203019 &     6.4&      K5V &    35.0&   400.       & & sat, ADI, H &  cc \\
      HIP105612    &       HD202732 &     6.3&      G5V &    32.8&   600.       & & sat, ADI, H &  \\
      HIP107684    &       HD207278 &     8.1&      G7V &    90.2&   100.       & Bin (new) & nonsat, ADI, H &  \\
      HIP108422    &       HD208233 &     6.9&     G9IV &    58.0&    30.       & Bin (known) & nonsat, ADI, H &  \\
 TYC\,8004-0083-1  &     CD-40-14901&     7.9&      G5V &    74.9&   100.       & & sat, ADI, H &  \\
      HIP114046    &       HD217897 &     3.6&      M2V &     3.3&  8000.       & & sat, ADI, H &  \\
 TYC\,9338-2016-1  &       HD220054 &     8.3&      G8V &    99.6&    30.       & & sat, ADI, H &  \\
 TYC\,9529-0340-1  &       CD-86-147&     7.6&     G8IV &    68.8&    30.       & & sat, ADI, H &  \\
 TYC\,9339-2158-1  &      CD-69-2101&     6.8&      K3V &    30.6&   300.       & & sat, ADI, H &  \\
 TYC\,6406-0180-1  &       HD221545 &     7.7&      K0V &    58.0&   200.       & & sat, ADI, H &  \\
      HIP116910    &       HD222575 &     7.8&      G8V &    63.7&   100.       & & sat, ADI, H &  \\
\noalign{\smallskip}\hline  \noalign{\smallskip}
\end{tabular}
\end{center}
\end{table*}


\section{Target Properties}

Based on a complete compilation of young, nearby stars, recently
identified in young co-moving groups and from systematic spectroscopic
surveys, we have selected a sample of stars according to: their
declination ($\delta \le 25^o$), their age ($\lesssim 200$~Myr), their
distance ($d \lesssim 100$~pc), their R-band brightness ($R \le 9.5$). In
addition, none of these stars had been observed in a high-contrast
imaging survey before.  Great care has been taken in the age selection
criteria based on different youth diagnostics (isochrones, lithium abundance,
H$\alpha$ emission, X-ray activity, stellar rotation, chromospheric
activity and kinematics). Close visual ($0.1-6.0~\!''$) and
spectroscopic binaries were rejected as they degrade the VLT/NaCo
detection performances and bias the astrophysical
interpretation. Among this sample, 86 stars were finally observed
during the large program. The main target properties (spectral type,
distance, age, H-band magnitude, galactic latitude, proper motion) are
reported in Tables~2 and 3. They are also shown in
Fig.~\ref{fig:sample_prop} together with the properties of the
complete statistical sample used by Vigan et al. (2014, in prep) and
Reggianni et al. (2014, in prep). A complete characterization of the
NaCo-LP observed sample and the archive sample, particularly the age
and distance determination, is detailed by Desidera et al. (2013,
submitted). As can be seen from Fig.~\ref{fig:sample_prop}, the core
of the NaCo-LP observed sample is mainly composed of young
(10--200~Myr), nearby solar-type FGK stars.

\section{Observations: Telescope and instrument}

\begin{table}[!t]

\caption{Observing campaigns with the ESO-program number, the
  observing mode (LP for Large-Program; OT for Open-Time; Vis for
  Visitor run and Ser for Service run), the starting night and the
  number of nights, the observing loss (technical and weather) and the
  number of observing sequences including single and multiple visits
  per target.}

\label{tab:obs}      
\begin{tabular}{llllll}     
\hline\hline       
ESO Program             & Mode      &  St. Night   & Night        & Loss       & Visit \\
                        &           &  (UT-date)      & (Nb)         & (\%)       & (Nb) \\
\hline
184.C-0157A           & LP-Vis    &  2009-11-21     & 3            & 20         & 23           \\
184.C-0157E           & LP-Vis    &  2010-02-16     & 3            & 0          & 27           \\
184.C-0157B           & LP-Vis    &  2010-06-14     & 2.5          & 70         & 11           \\
184.C-0157F           & LP-Vis    &  2010-07-29     & 2            & 33         & 18           \\
184.C-0157C           & LP-Ser    &  -              & 1.5          & 0          & 15           \\
184.C-0157D           & LP-Ser    &  -              & 3.3          & 0          & 33           \\
089.C-0137A           & OT-Ser    &  -              & 0.7          & 0          & 6            \\
090.C-0252A           & OT-Ser    &  -              & 0.5          & 0          & 4            \\
\hline                  
Total                   & -         & -               & 16.5         &            & 137          \\ 
\hline                  
\end{tabular}
\end{table}

\begin{figure*}[t]
\hspace{0.2cm}
\includegraphics[width=5.6cm]{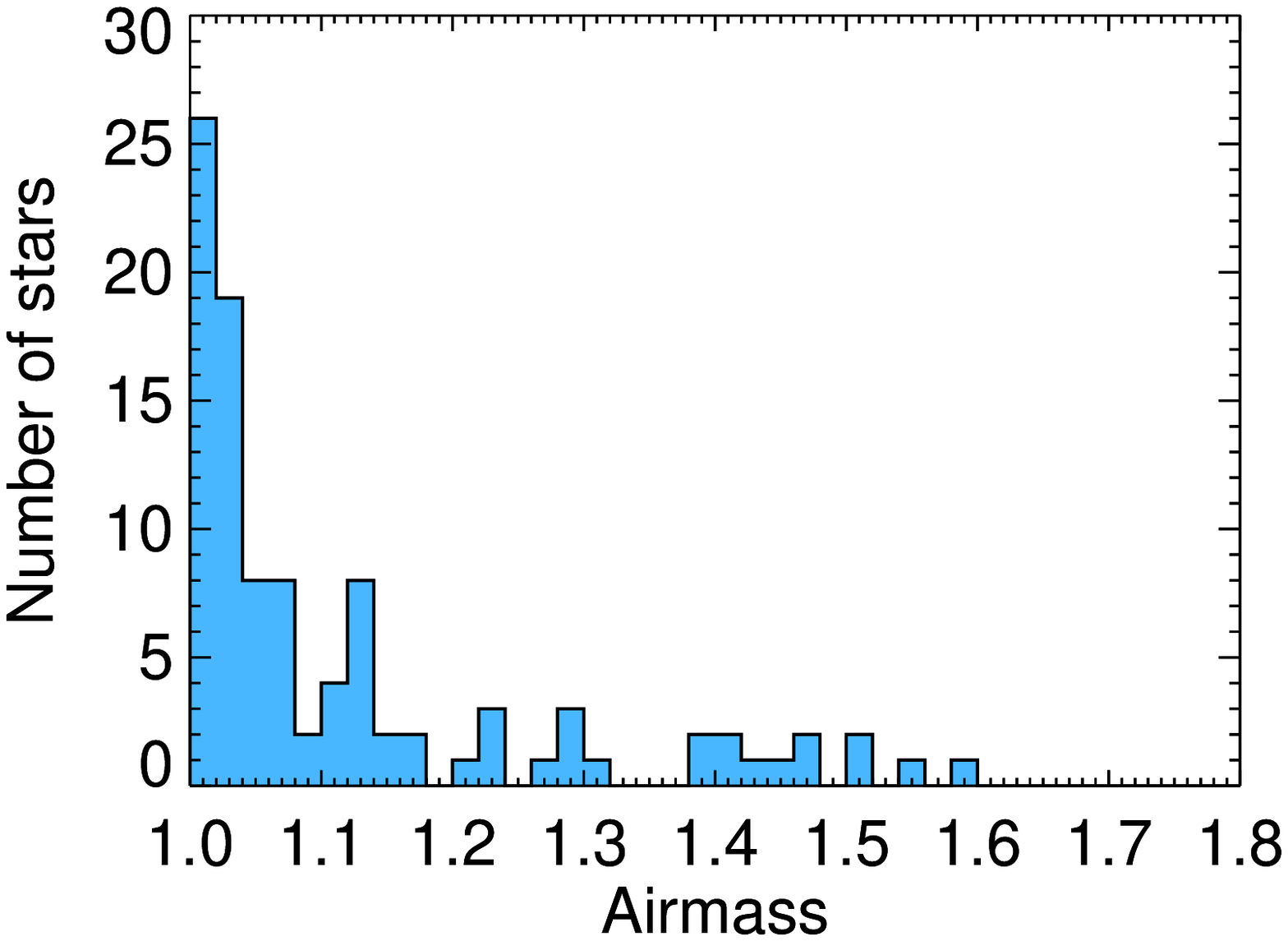}
\includegraphics[width=5.6cm]{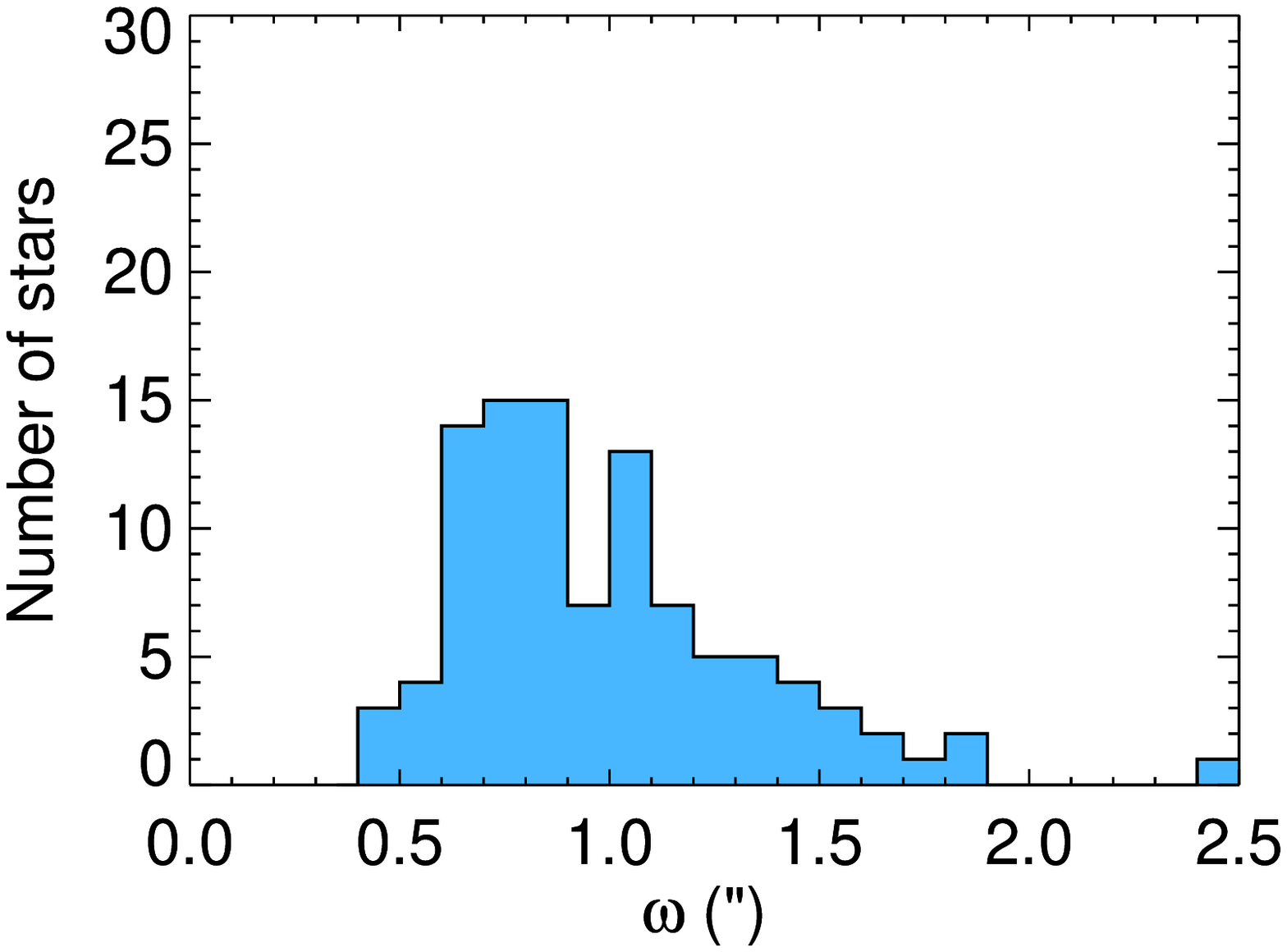}
\includegraphics[width=5.6cm]{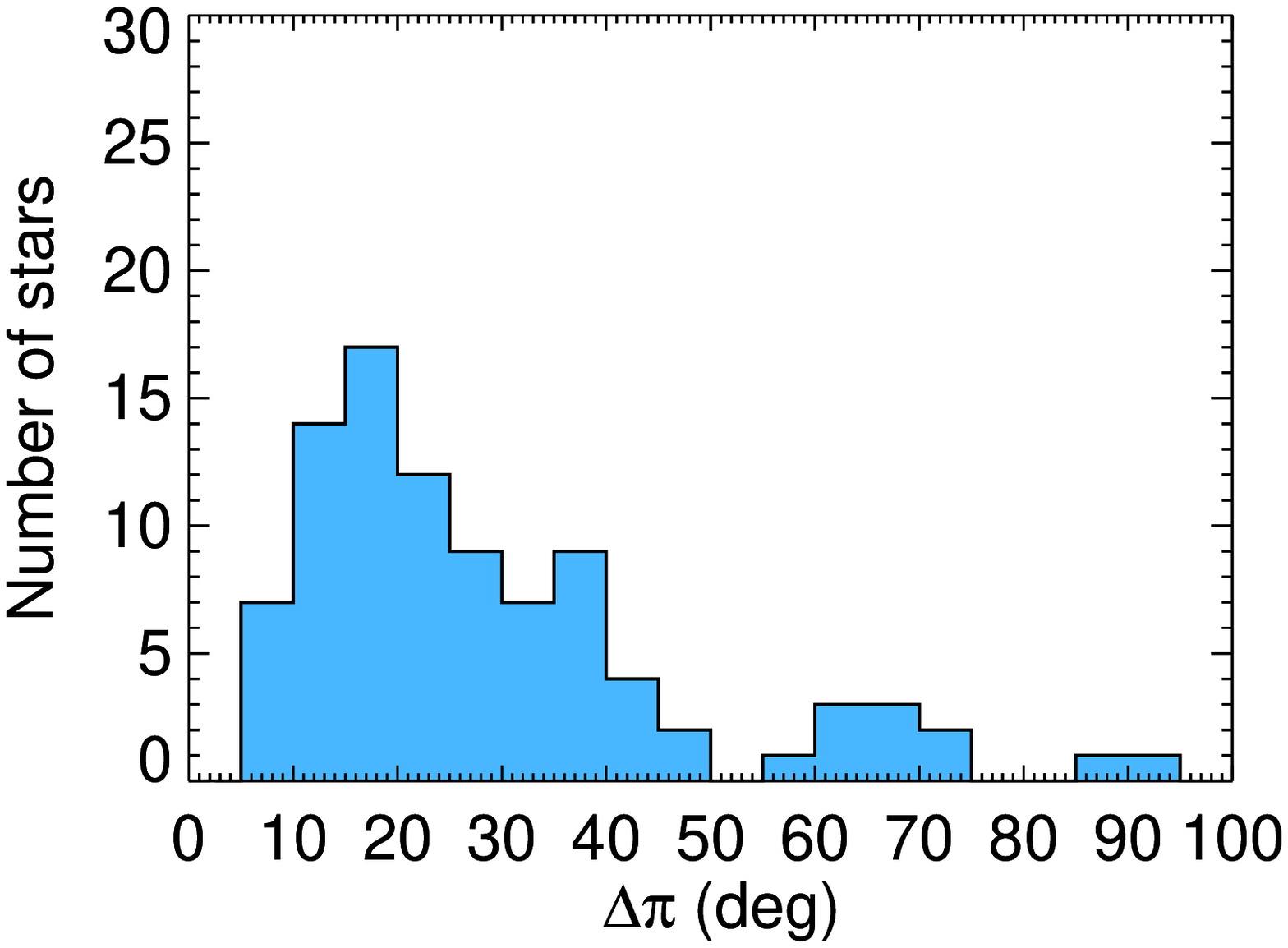}\\
\includegraphics[width=5.6cm]{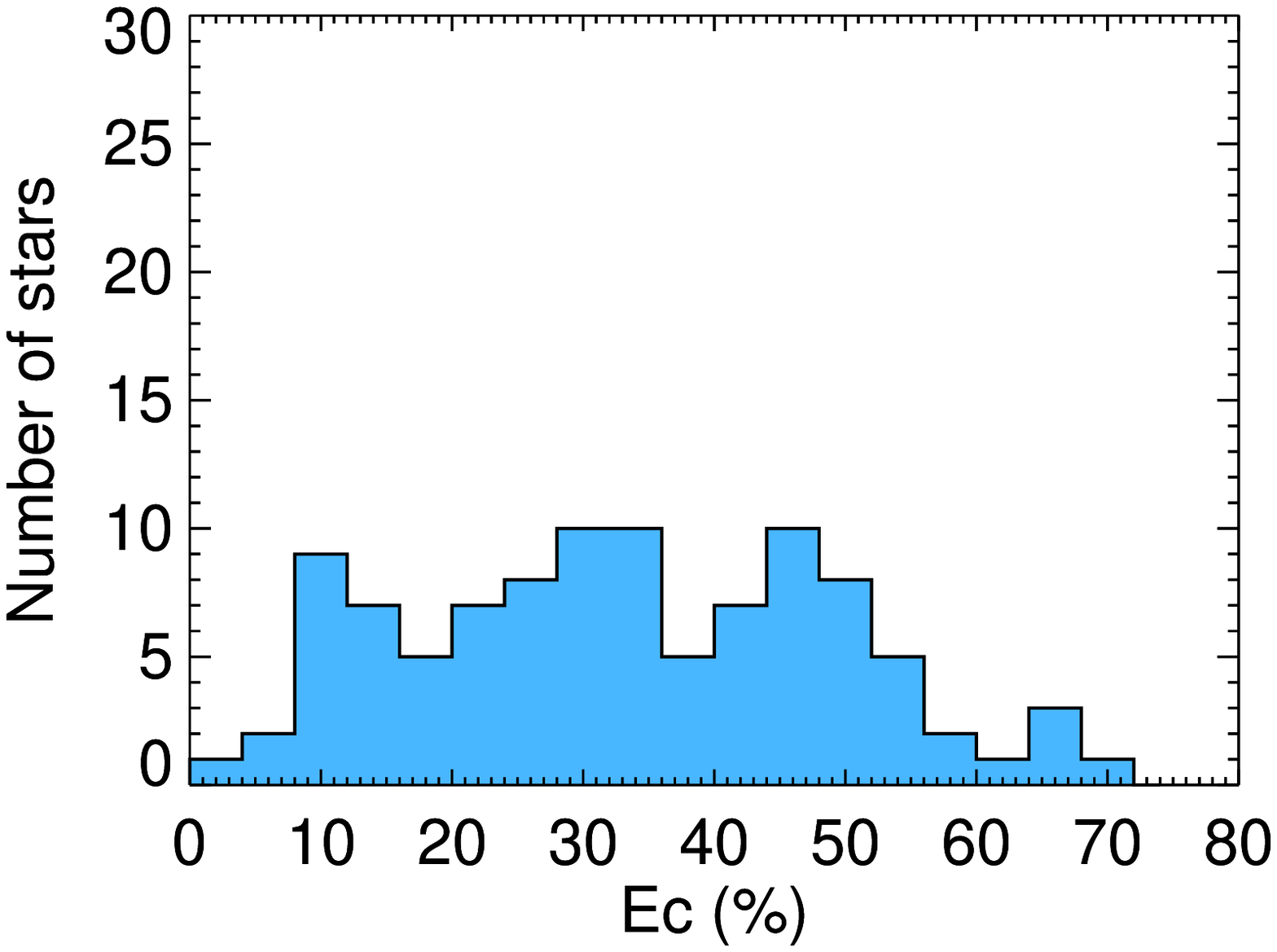}
\includegraphics[width=5.6cm]{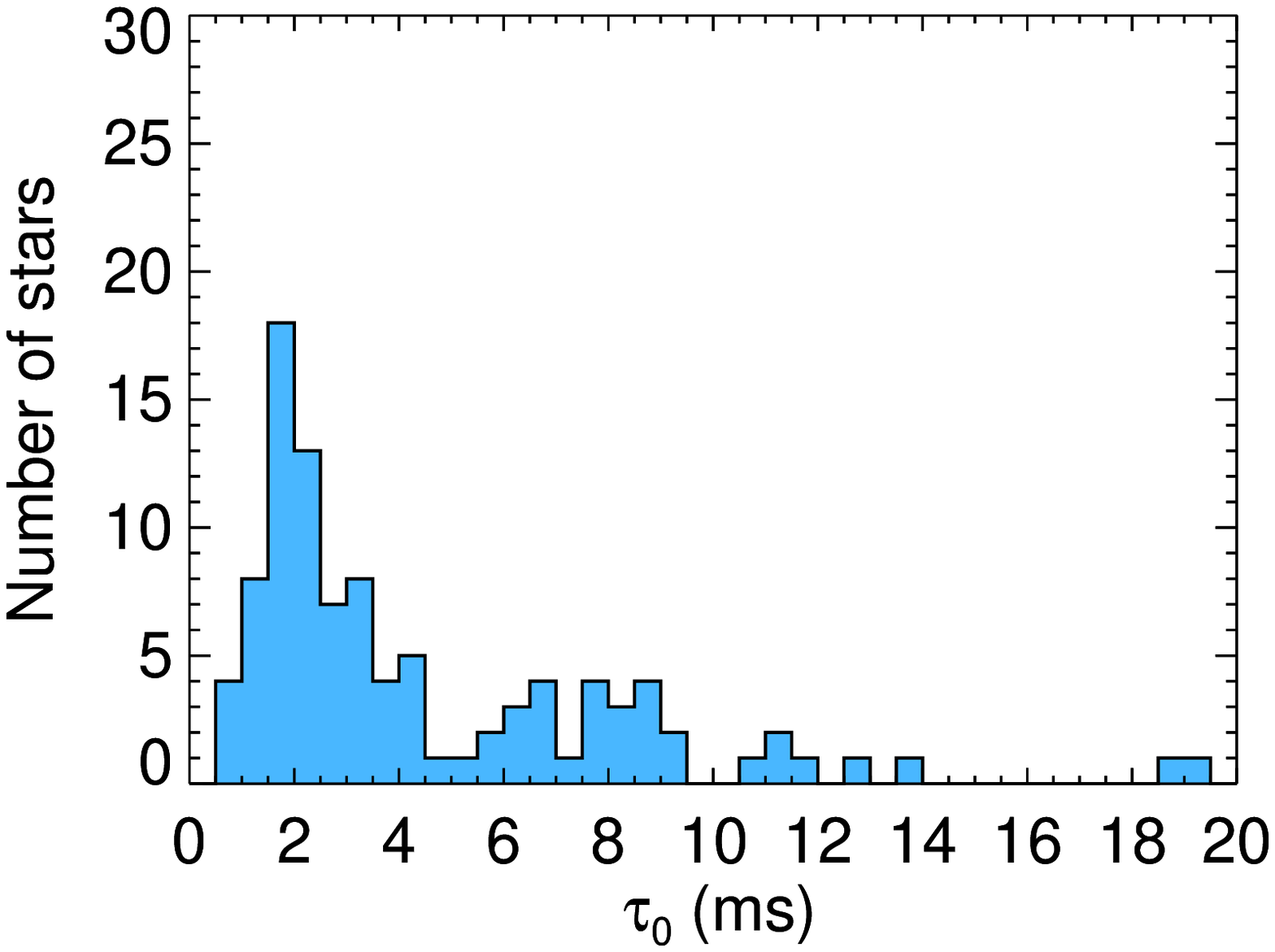}
\begin{centering}

\caption{Histograms summarizing the observing conditions of the
  NaCo-LP campaigns: airmass, DIMM seeing ($\omega$), parallactic angle
  variation ($\Delta\pi$), coherent energy (Ec) and coherent time
  ($\tau_0$).}

\label{fig:obscond}
\end{centering}
\end{figure*}

We used the NaCo high contrast Adaptive Optics (AO) imager of the
VLT-UT4. NaCo is equipped with the NAOS AO system (Rousset et
al. 2002), and the near infrared imaging camera CONICA (Lenzen et
al. 2002). The observations were obtained during various observing
runs spread between end 2009 and 2013 in visitor and service
(queue-observing) modes. The summary of the observing runs is reported
in Table~4.  The NaCo-LP represents a total of 16.5 observing nights,
10.5 nights obtained in visitor mode and 6 nights in service.

To achieve high contrasts, we used the angular differential imaging
(ADI) on pupil-stabilized mode of NaCo. A classical Lyot-coronagraph
with a diameter of $0.7~\!''$ was used during the first visitor run
but then replaced by saturated imaging as the NaCo PSF was
unexpectedly drifting with time owing to a technical problem with the
instrument. For accurate astrometry, a single observing set-up was
used, corresponding to the combined use of the $H$-band filter with
the S13 camera (13.25~mas/pix). The time of the observations were
chosen to maximize the field rotation. Typical exposure times of
1--10s were used to saturate the PSF core by a factor 100 (a few
pixels in radius) to improve the dynamic range of our images. The NaCo
detector cube mode was in addition used to register each individual
frames to optimize the final image selection in post-processing. The
typical observing sequence was composed of a total of 10--15 cubes of
10--120 frames, i.e a total integration time of 35--40~min for an
observing sequence of 1--1.5~hrs on target. The parallactic angles
variations are reported in Fig.~\ref{fig:obscond} together with the
airmass, coherent energy and coherent time measured by NaCo and the
seeing measured by the DIMM seeing monitor at VLT. Non-saturated PSFs
were acquired in ADI using a neutral density filter at the beginning
of each observing sequence to monitor the image quality. They also
served for the calibration of the relative photometric and astrometric
measurements.

\section{Data reduction and analysis}

\subsection{Cosmetics and data processing}

\begin{figure}[t]
\hspace{0.2cm}
\includegraphics[width=7cm]{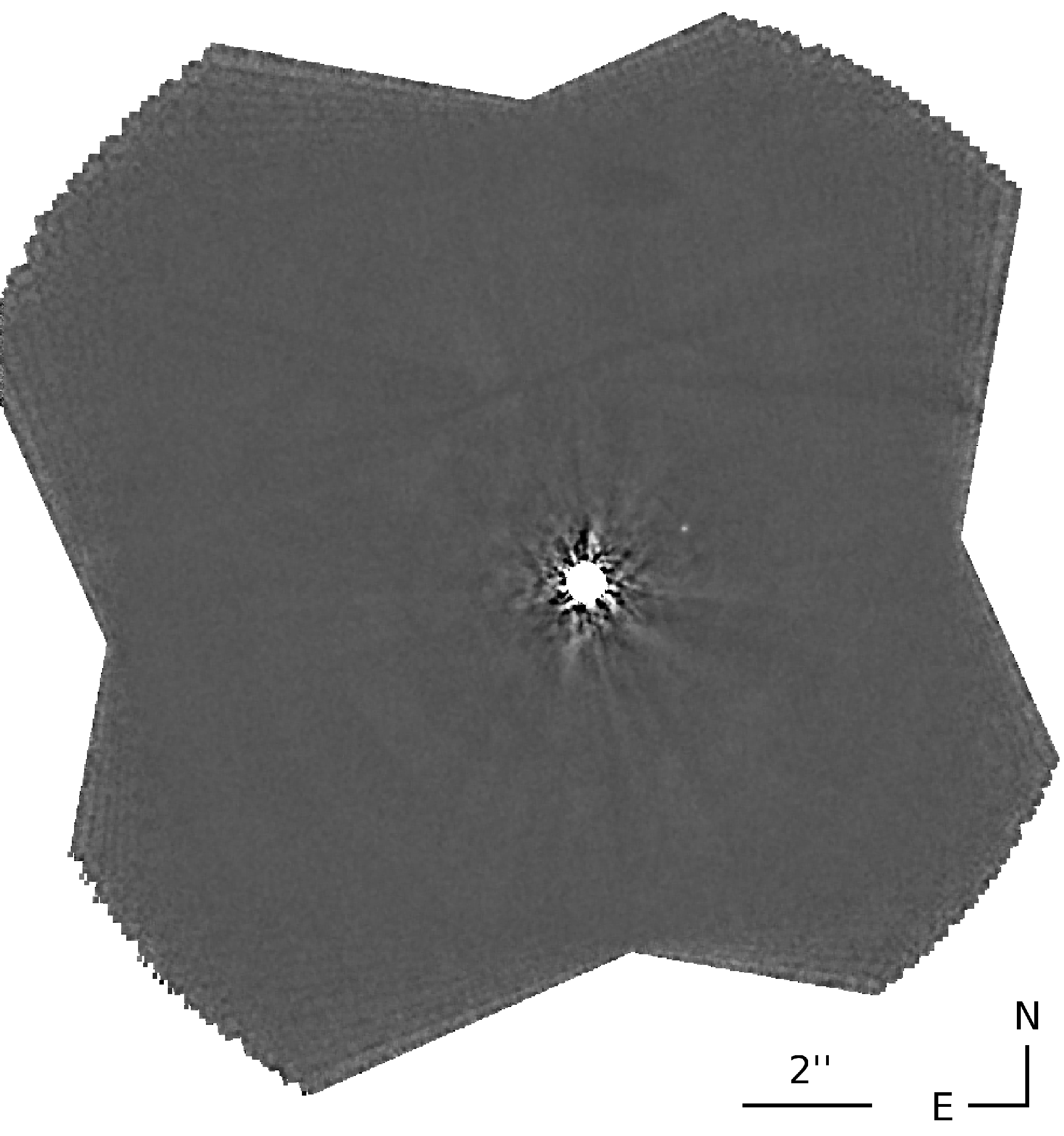}
\begin{centering}

\caption{VLT/NACO ADI observation in H-band combined of the young star
  TYC\,7617-0549-1 (K0V, 76.4~pc and 30~Myr). A faint ($\Delta
  H=12.7$~mag) candidate, resolved at $1.8~\!''$, has been finally
  identified as a background contaminant (see Fig.~\ref{fig:ppm}). }

\label{fig:image}
\end{centering}
\end{figure}
\begin{table}[t]
\begin{centering}
\caption{Mean plate scale and true north orientation for each
  observing run of the NaCo-LP. }
\label{tab:calastro}      
\begin{tabular}{llll}     
\hline\hline       
UT Date     & Platescale        & True north                            & Calibrator  \\
            & (mas)             & (deg)                                 &             \\
\hline
2009-11-23  &   $13.22\pm0.02$  &                      $-0.17\pm0.03$   & $\theta_1$ Ori C\\             
2010-02-15  &   $13.21\pm0.02$  &                      $-0.33\pm0.03$   & IDS1307\\             
2010-02-18  &   $13.21\pm0.02$  &                      $-0.33\pm0.03$   & $\theta_1$ Ori C\\  
2010-06-16  &   $13.21\pm0.02$  &                      $-0.53\pm0.03$   & IDS1307\\  
2010-07-30  &   $13.21\pm0.02$  &                      $-0.47\pm0.03$   & IDS1307\\  
2010-12-30  &   $13.21\pm0.02$  &                      $-0.47\pm0.03$   & $\theta_1$ Ori C\\ 
2011-01-30  &   $13.21\pm0.02$  &                      $-0.49\pm0.03$   & $\theta_1$ Ori C\\      
2011-05-11  &   $13.21\pm0.02$  &                      $-0.52\pm0.03$   & IDS1307\\  
2011-07-02  &   $13.21\pm0.02$  &                      $-0.55\pm0.03$   & IDS1307\\  
2012-01-02  &   $13.22\pm0.02$  &                      $-0.59\pm0.03$   & IDS1307\\ 
2012-01-02  &   $13.22\pm0.02$  &                      $-0.59\pm0.03$   & $\theta_1$ Ori C\\          
\hline                  
\end{tabular}
\end{centering}
\end{table}

Three independent pipelines were used to reduce and analyse the ADI
data in order to optimize the PSF-subtraction and the detection
performances and to check the consistency of the results in terms of
astrometry and photometry. These pipelines are described for: the
LAM-ADI pipeline by Vigan et al. (2012), the IPAG-ADI pipeline by
Chauvin et al. (2012) and the Padova-ADI pipeline by Esposito et
al. (2013). Each pipeline processed the data in a similar way for the
first cosmetics steps of flat-fielding, bad and hot-pixels removal,
and sky-subtraction. To determine the central star position, as for
the frames recentering, a Moffat fitting of the non-saturated part of
the stellar PSF wing (with a similar threshold) was used. Finally, an
encircled energy criteria was considered for the rejection of
open-loop and poorly-corrected frames to compute a final mastercube
together with the correspoding parallactic angle variation. The main
differences between the pipelines mostly reside in the various ADI
algorithms applied (cADI and sADI, see Marois et al. 2006; LOCI, see
Lafreni\`ere et al. 2007) and in the parameters setup. Consistent
results within 0.1--0.2~mag in photometry (candidate photometry and
detection limits) and 0.2--0.3 pixels in astrometry were found between
the different pipelines for a serie of targets used as test
cases. Non-saturated PSFs were similarly reduced without
PSF-subtraction.

The results presented in this final analysis have been obtained with
the LAM-ADI pipeline using LOCI, considering optimization regions of
$N_A = 300\times FWHM$ at less than $3~\!''$ and $N_A = 3000\times
FWHM$ at more than $3~\!''$, the radial to azimuthal width ratio
$g=1$, the radial width $\Delta r=2\times FWHM$ and a separation
criteria of $0.75\times FWHM$. The binning of the data was tuned to
apply LOCI on a final mastercube reduced to $\sim350$ frames. An
illustration of the final LOCI processed image of the young star
TYC\,7617-0549-1 (K0V, 76.4~pc and 30~Myr) is shown in
Fig.~\ref{fig:image}.

\subsection{Relative astrometry and photometry}

The relative position and flux of all candidates was
determined using Moffat fitting and aperture photometry
corrected from the ADI flux loss. This first order analysis
  was sufficient to assess the candidate proper motion and nature as
  described in sub-section 5.2. For the most interesting cases
(like HD\,8049), the injection of fake planets at the
location of the candidate signal was done to properly take into
account any local astrometric and photometric biases induced
by the ADI-processing described in Chauvin et al. (2012).

To finally calibrate the relative astrometric position of the detected
candidates to the primary star, we used the $\theta_1$ Ori C field
observed with HST by McCaughrean \& Stauffer (1994) (with the same set
of stars TCC058, 057, 054, 034 and 026) as a primary calibrator. The
astrometric binary IDS\,13022N0107 (van Dessel \& Sinachopoulos 1993)
was then used as a secondary calibrator when the $\theta_1$ Ori C
field was not observable and then recalibrated on the $\theta_1$ Ori C
field when both were observable. Both fields were observed in standard
field-stabilized mode and reduced (cosmetics, flat-fielding, bad and
hot-pixels removal, sky-subtraction and recentering) using the
\textit{Eclipse}\footnote{http://www.eso.org/projects/aot/eclipse/}
reduction software developed by Devillar (1997).  Finally, for ADI
data, the NaCo rotator offset at the start of each ADI sequence was
also calibrated and taken into account as described by Chauvin et
al. (2012). The results of the platescale and true north orientation
determinations are given in Table~\ref{tab:calastro}.

The throughput of the NaCo neutral density filter was recalibrated on
sky using two different datasets taken for the star TYC\,9162-0698
during our February 2010 visitor run. Using aperture photometry on the
data taken with and without the neutral density, we derived a
transmission factor of $1.19\pm0.05\%$ with the H-band filter. This
result is consistent with the one derived by Bonnefoy et
al. (2013a) and was used to calibrate the candidate photometry
and the detection limits using the non-saturated sequence of the
primary star with the neutral density filter as a photometric reference.

\begin{figure}[t]
\hspace{0.2cm}
\includegraphics[width=9cm]{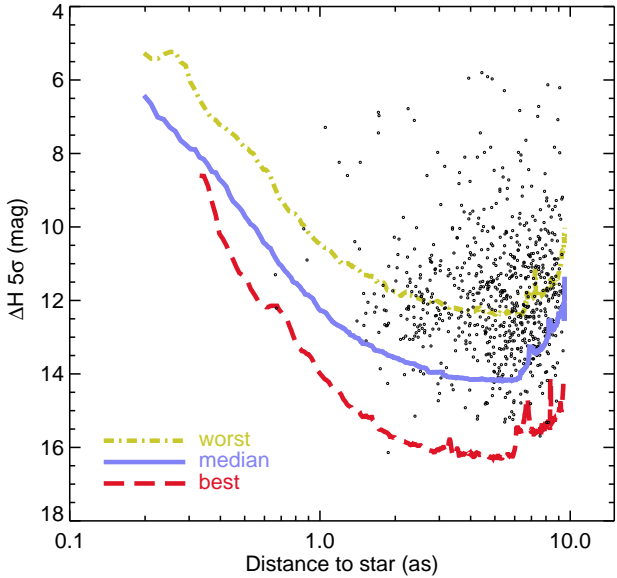}
\begin{centering}
\caption{VLT/NACO deep ADI 5$\sigma$ detection limits in H-band combined with
  the S13 camera. The worst, median and best detection limits
  are shown with all the candidates detected. Separations of less than
  0.1--0.2$~\!''$ are generally saturated.}
\label{fig:detlim}
\end{centering}
\end{figure}

\subsection{Detection limits determination}

A pixel-to-pixel noise map of each observation was estimated
  within a box of 5$\times$5 pixels sliding from the star to the limit
  of the NACO field of view. To correct for the flux loss related to
the ADI processing, fake planets were regularly injected every
20~pixels in radius at 10 different position angles for separations
smaller than $3~\!''$. At more than $3~\!''$, fake planets were
injected every 50~pixels at 4 different position angles. The final
flux loss was computed with the azimuthal average of the flux losses
of fake planets at same radii. The final detection limits at $5\sigma$
were then obtained using the pixel-to-pixel noise map divided
by the flux loss and normalized by the relative calibration with the
primary star (considering the different exposure times and the neutral
density). LOCI processing leads to residuals whose distribution
closely resembles a Gaussian (Lafreni\`ere et al. 2007), therefore a
$5\sigma$ threshold is thus adequate for estimating detection
performances. The best, worst and median detection limits of the
survey are reported in Fig.~\ref{fig:detlim}.

\section{Results}

A total of 86 sources were observed. 16 stars were resolved as
binaries, including HD\,8049 with a newly discovered white dwarf
companion. 10 binaries were simply observed in non-saturated ADI
imaging to directly derive their relative astrometry and
photometry. 76 stars were observed in saturated high-contrast ADI to
search for faint substellar companions. In the following sub-sections,
we describe the properties of the new stellar multiple systems, the
status of the detected candidates in saturated ADI, the
characteristics of the white dwarf companion around HD\,8049, finally
the fine analysis of the thin debris-disk around HD\,61005.


\subsection{New stellar close multiple systems}

Despite our sample selection to reject close ($0.1-6.0~\!''$)
binaries, 16 stars were resolved as multiple. Three systems were
already known, HIP\,108422\,AB (Chauvin et al. 2003),
TYC\,7835-2569-1\,AB (Brandner et al. 1996) and TYC\,6786-0811-1
(K\"ohler et al. 2000), and went through our sample selection process
by mistake. TYC\,8484-1507-1 is actually also a known
$\sim8.6~\!''$ binary that was resolved by 2MASS, not rejected during
our sample selection, but resolved in the NaCo FoV despite its large
separation. Then, in the case of HD\,8049, the faint comoving
companion turned out to be a white dwarf. Its characteristics are
briefly described in Subsection~\ref{subsec:wd}). At the end, a total
of 11 new close multiple systems were resolved. All of them
  were observed in non-saturated ADI to derive their position and
  $H$-band photometry relative to the primary star (see Table~6). The
  visual binaries HIP\,108422\,AB, TYC\,7835-2569-1\,AB and
  TYC\,6786-0811-1 are confirmed as physically bound. Deep ADI
  observations were obtained in addition for six binaries
  (TYC\,0603-0461-1, TYC\,7835-2569-1, HD\,8049, TYC\,8927-3620-1,
  HIP\,80290 and TYC\,8989-0583-1).

\subsection{Companion candidates}

\begin{table}[t]

\caption{Relative positions and $H$-band contrast of the new binaries
  resolved during the NaCo-LP. Epochs of observation are reported in
  Tables~7, 8, 9, 10 and 11.}

\label{tab:new_binaries}      
\begin{tabular}{llll}     
\hline\hline\noalign{\smallskip}       
Name                    &  $\Delta$    & PA               & $\Delta H$            \\ 
                        &  (mas)       & (deg)            & (mag)                 \\
\noalign{\smallskip}\hline\noalign{\smallskip}
HIP\,8038$^a$           & $437\pm7$    & $273.8\pm0.9$    & $2.5\pm0.2$           \\ 
HIP\,80290              & $3340\pm4$   & $257.5\pm0.2$    & $1.9\pm0.2$   \\
HIP\,94235              & $506\pm7$     & $150.6\pm0.8$   &  $3.8\pm0.3$                     \\ 
HIP\,107684             & $326\pm7$    & $270.2\pm1.2$    & $2.8\pm0.3$           \\ 
HD\,199058              & $471\pm7$    & $282.9\pm0.8$    & $2.5\pm0.2$           \\ 
TYC\,0603-0461-1        & $74\pm14$    &  $74.2\pm5.1$    &  $0.1\pm0.4$                       \\ 
TYC\,8927-3620-1$^b$        & $87\pm14$     & $296.8\pm4.35$   & $0.5\pm0.4$           \\
TYC\,8989-0583-1        &  $2584\pm8$  & $169.8\pm0.15$   & $2.7\pm0.2$           \\
TYC\,9010-1272-1        &$262\pm8$     & $238.0\pm1.44$   & $1.0\pm0.3$           \\
TYC\,9181-0466-1        & $1891\pm7$    & $123.4\pm0.2$   & $1.5\pm0.2$           \\
TYC\,9231-1566-1        & $1975\pm7$    & $145.5\pm0.2$   & $3.0\pm0.2$           \\

\noalign{\smallskip}\hline                  
\end{tabular}
\begin{list}{}{}
\item[\scriptsize{(a):}] \scriptsize{Known binary separated by 15.0~arcsec and $\Delta V=2.0$~mag.}
\item[\scriptsize{(b):}] \scriptsize{Third component resolved by 2MASS at $\sim4.8~\!''$ and $\Delta K=0.7$~mag.}
\end{list}
\end{table}

\begin{figure}[t]
\includegraphics[width=9cm]{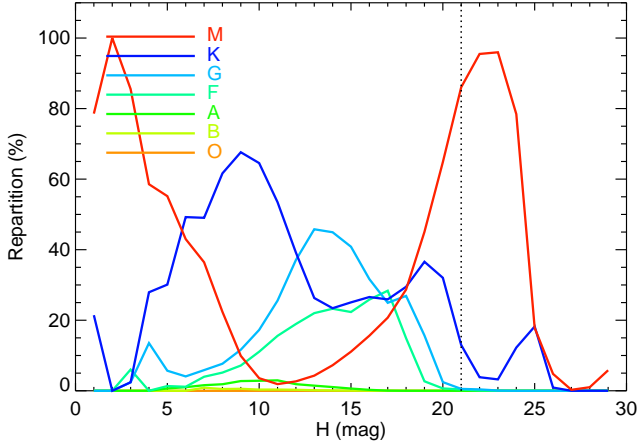}
\begin{centering}

\caption{Expected spectral type distribution of field stars
    from the Besan\c{c}on galactic population model observed during
    the NaCo-LP. The FoV, the typical magnitude limit of the NaCo-LP
  ($H_{\rm{lim}}=21$~mag) and the galactic coordinates of all targets
  were considered.  The predicted repartion is given as a function of
  the spectral type and the apparent magnitude in $H$-band.}

\label{fig:conta}
\end{centering}
\end{figure}

Among the 76 stars observed in ADI, one companion candidate or more
were detected for 43 targets (see Tables~2 and 3). More than $700$
candidates were detected, 90\% of them in six very crowded field (see
Fig.~\ref{fig:detlim}). The galactic contamination rate of the NaCo-LP
fields by at least one background source predicted by the Besan\c{c}on
galactic population model (Robin et al. 2003) is equal to $51\%$, in
reasonable agreement with the $56\%$ (43 systems with at least one
candidate for the 76 observed). The model uses as input the NaCo field
of view, the typical magnitude limit of the NaCo-LP
($H_{\rm{lim}}=21$~mag) and the galactic coordinates of all
targets. The repartition of these galactic contaminants is given in
Fig.~\ref{fig:conta}. Solar-system and extra-galactic contaminants are
expected to be significantly less frequent. Moreover, solar system
contaminants smear during a 1-hr observing sequence and extra-galactic
contaminants are mainly extended background galaxies resolved by
NaCo. The most important population of contaminants that can mimic the
apparent flux of giant planet or brown dwarf companions bound to the
star are M dwarfs with typical $H=20-22$~mag apparent magnitudes.

To identify their nature, we relied on the follow-up observations at
additional epochs to distinguish comoving companions from stationary
background stars. The candidates were ranked by priority as a function
of their predicted masses (higher priority to lower masses),
projected physical separations (assuming they would be bound;
higher priority to closer candidates) and predicted false
alarm probabilities using the Besan\c{c}on galactic population model
(Robin et al. 2003) to guide the follow-up strategy. Follow-up
observations with a second epoch were obtained for 29 targets,
including the Moth system (HD\,61005) characterized during
dedicated follow-up observations. The amplitude of stellar proper
motion (larger than 30~mas/yr for $80\%$ of the NaCo-LP target)
enabled a rapid identification over 1~yr interval (see
Fig.~\ref{fig:sample_prop}, \textit{Bottom-Middle}).

\begin{figure}[t]
\hspace{0.2cm}
\includegraphics[width=9cm]{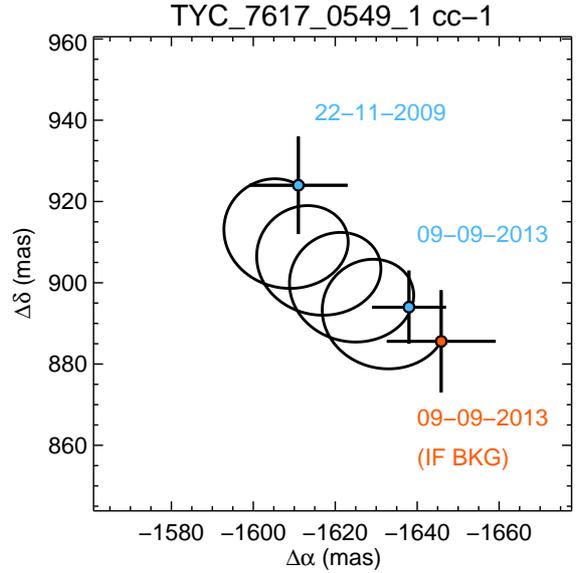}
\begin{centering}
\caption{VLT/NaCo measurements (filled circles with uncertainties) of
  the offset positions of the companion candidate to TYC\,7617-0549-1
  (see Fig.~\ref{fig:image}). The expected variation of offset positions, if the
  candidate is a background object, is shown (curved line). The
  variation is estimated based on the parallactic and proper motions
  of the primary star, as well as the initial offset position of the
  companion candidate from TYC\,7617-0549-1. The companion candidate is
  clearly identified here as a stationary background contaminant.}

\label{fig:ppm}
\end{centering}
\end{figure}

For the 29 systems with at least 2-epochs observations (including the
Moth system), we used a $\chi^2$ probability test of $2\times
N_{epochs}$ degrees of freedom (corresponding to the measurements:
separations in the $\Delta\alpha$ and $\Delta\delta$ directions for
the number $N_{epochs}$ of epochs). This test takes into account the
uncertainties in the relative positions measured at each epoch and the
uncertainties in the primary proper motion and parallax (or
distance). Fig.~\ref{fig:ppm} gives an illustration of a
($\Delta\alpha$, $\Delta\delta$) diagram that was used to identify a
stationary background contaminant around TYC\,7617-0549-1.  A status
has been assigned to each candidate as background contaminant (B;
P$_{\rm{comoving}, \chi^2}<1$\% and with a relative motion compatible
with a background source), comoving (C; P$_{\rm{BKG}, \chi^2}<1$\%)
and with the relative motion compatible with a comoving companion) and
undefined (U) when observed at only one epoch or when not satisfying
the first two classifications.

Only one comoving companion was identified, the white dwarf companion
around HD\,8049 described here after. Among the 28 other follow-up
fields, 10 fields have been completely characterized and 18 partially
owing to detection limits variation from one epoch to another. 14
fields still require second epoch observations. The status of all the
candidates is given in Tables~7, 8, 9, 10 and 11.

\subsection{A white dwarf companion around HD\,8049}
\label{subsec:wd}      

The only comoving companion identified in this survey, with a
preliminary predicted mass of 35~M$_{\rm{Jup}}$, was discovered around the star
HD\,8049 (K2, 33.6~pc). The star had a predicted age of 90--400~Myr
from its rotational period, H\&K emission and X-ray emission. Thanks to
the high proper motion of the central star ($\mu_{\alpha} =
65.99\pm1.18$ mas/yr and $\mu_{\delta} = −240.99\pm0.98$ mas/yr), a
$\chi^2$ probability test on $\Delta\alpha$ and $\Delta\delta$ with
respect to the star at two epochs rejected the possibility (at 99\%
certainty) that the object was a background source.  Further analysis
using archived data, radial velocity observations spanning a time
range of $\sim30$~yr, U-band imaging with EFOSC and near-infrared
spectroscopy of the comoving companion with VLT/SINFONI finally
revealed that the companion was actually a white dwarf (WD) with
temperature $T_{\rm{eff}}=18800\pm2100$~K and mass M$_{\rm{WD}} =
0.56\pm0.08~$M$_{\odot}$. 

This astrophysical false positive revealed that the system age was
much older than initially thought. The age diagnostics have likely been
affected as the central star has been probably rejuvenated by the
accretion of some amount of mass and angular momentum at the time of
mass loss from the WD progenitor. A complete analysis of the system
(evolution, kinematics) by Zurlo et al. (2013) actually reveals that
the resulting age of the system is about 3--6~Gyr.

\subsection{The Moth resolved as a thin debris-disk}

In the course of the survey, the emblematic star HD\,61005 (G8V,
90~Myr, 34.5~pc), known to host “The Moth” debris disk (Hines et
al. 2007), was observed. The NaCo $H$-band image remarkably resolves
the disk component as a distinct narrow ring at inclination of
$i=84.3\pm1.0^o$, with a semimajor-axis of $a =61.25\pm0.85$~AU and an
eccentricity of $e = 0.045\pm0.015$. The observations also revealed
that the the ring center is offset from the star by at least
$2.75\pm0.85$~AU indicating a possibly dynamical perturbation by a
planetary companion that perturbs the remnant planetesimal belt. The
observations and the detailed disk modeling was published by Buenzli
et al. (2010). Subsequent observations did not reveal any giant planet
companions. Three other stars of our sample are known to host
debris-disks: HIP\,11360 (HD\,15115; Kalas et al. 2007, Rodigas et
al. 2012), HIP\,99273 (HD\,191089; Churcher et al. 2011) and
HIP\,76829 (HD\,139664; Kalas et al. 2006). No clear detection was
obtained with our ADI analysis.

\section{Survey's statistical analysis}

\begin{figure*}[t]
\includegraphics[width=9cm]{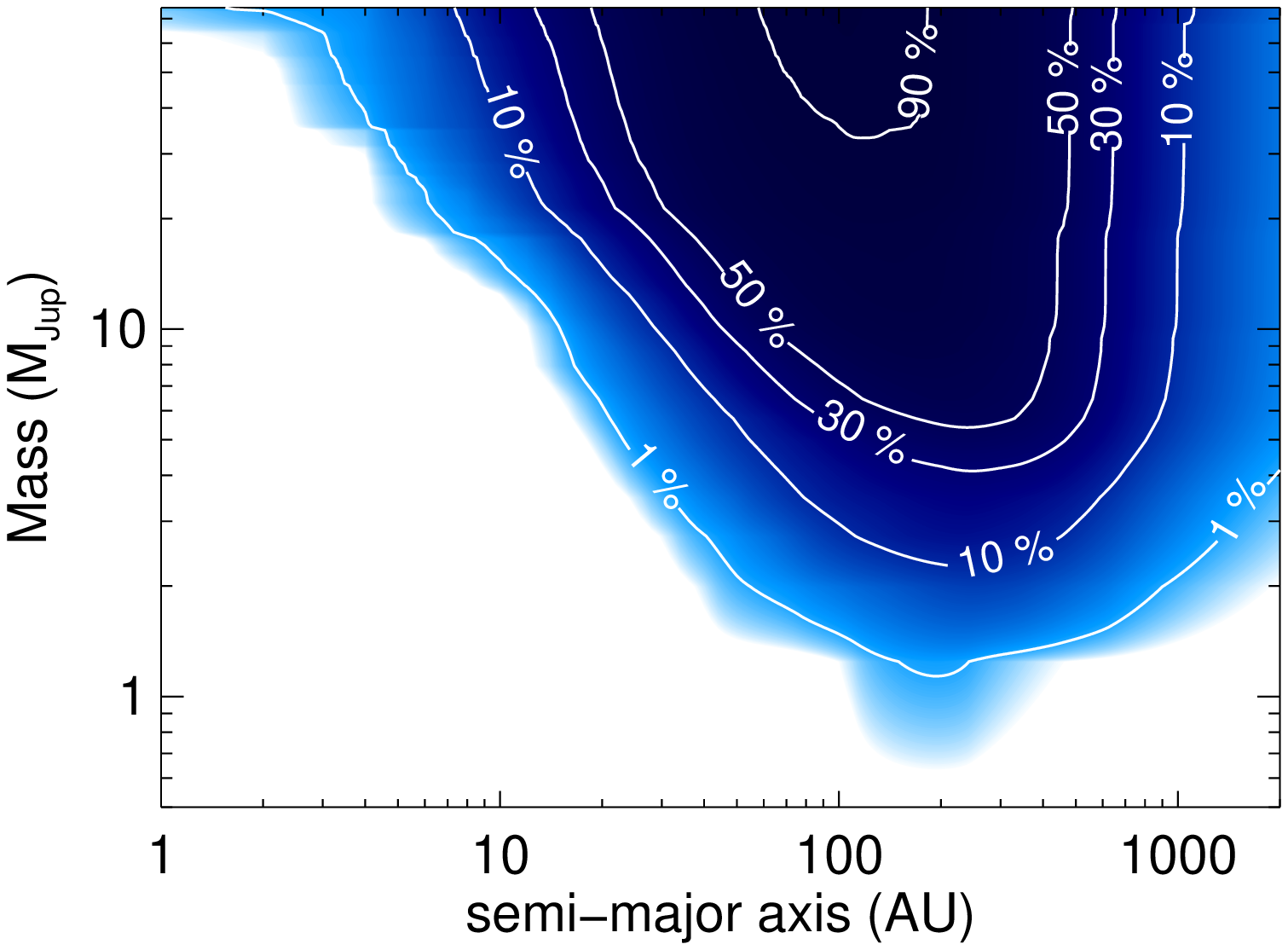}
\includegraphics[width=9cm]{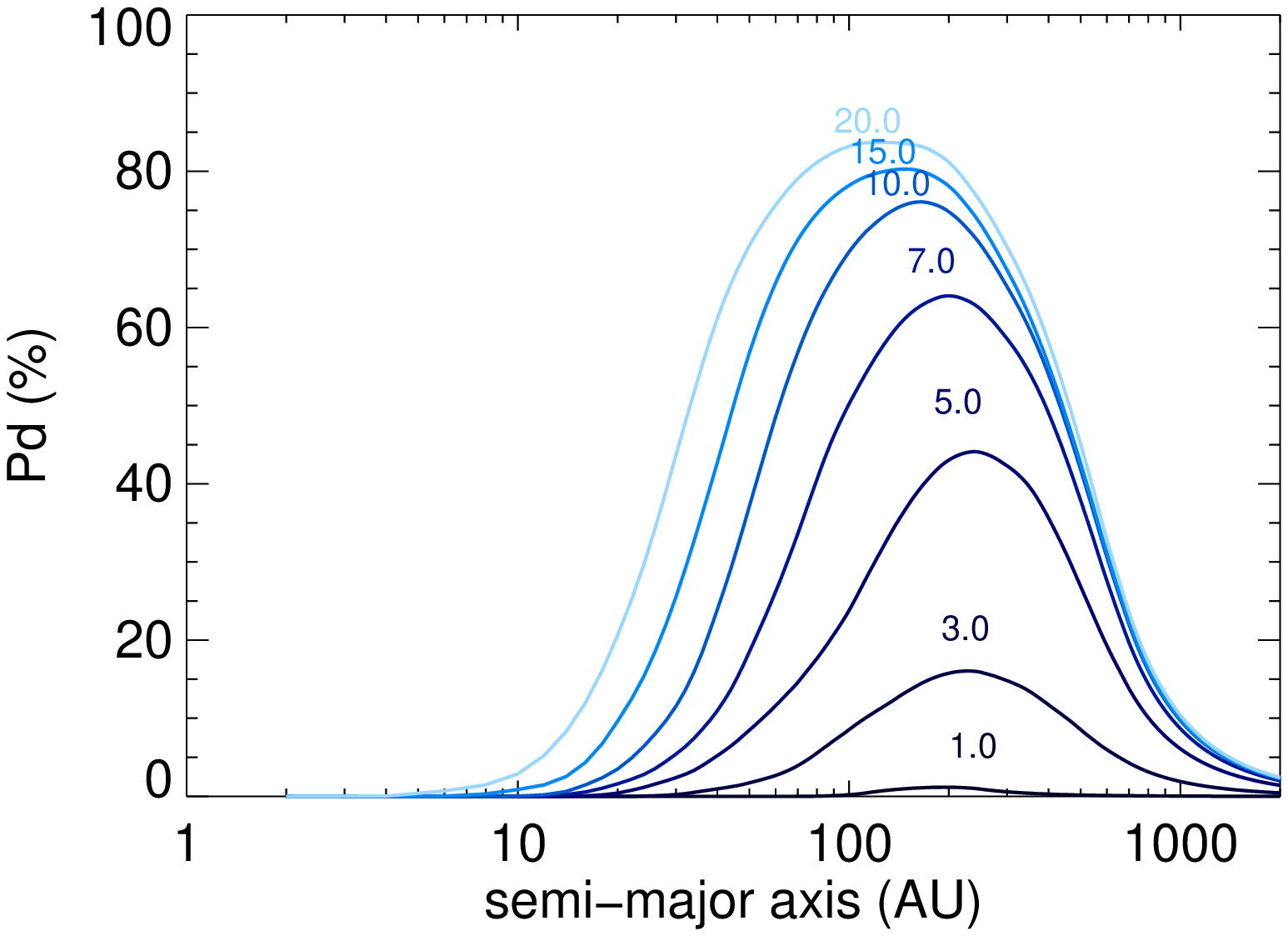}
\begin{centering}
\caption{\textit{Left:} NaCo-LP mean detection probability map
  ($<p_j>$) as a function of the mass and semi-major
  axis. \textit{Right:} Mean probability curves for different masses (1, 3,
  5, 7, 10, 15 and 20~M$_{\rm{Jup}}$) as a function of the semi-major
  axis.}
\label{fig:detprob}
\end{centering}
\end{figure*}

\subsection{Sample definition}

To define a meaningful sample for the statistical analysis of the
survey, we first removed from the sample of 76 stars observed in ADI
all visual and spectroscopic binaries, including the six visual
multiple systems observed in that mode (TYC\,0603-0461-1,
TYC\,7835-2569-1, HD\,8049, HIP\,8290, TYC\,8927-3620-1 and
TYC\,8989-0583-1) and 7 new spectroscopic binaries unknown at the time
of our sample selection. We have then selected two sub-samples:

\begin{itemize}
\item the \textit{full-stat} sample of 63 stars including all single
  stars observed in ADI with detection sensitivities down to planetary
  masses for physical separations ranging from 10 to 2000~AU. The
  status of all the candidates detected in these fields have however
  not been fully completed, although a large majority are expected to
  be stationary background contaminants. This sample gives an
  estimation of the ultimate performances of the survey interms of
  masses and physical separations when the candidate status
  identification will be complete, probably with SPHERE in the
  forthcoming years.

\item the \textit{complete-stat} sample of 51 stars has been
  restreined to all systems for which the candidate status
  identification was complete up-to 300~AU, including cases with no
  companion candidates detected or with companion candidates properly
  identified thanks to our follow-up observations as stationary
  background sources or comoving companions. In the case of follow-up
  observations with variable detection performances from one epoch to
  another (therefore with possible undefined faint sources due to the
  lack of redetection), only the worst detection limit was
  considered. These selection criteria offered us at the end a
  meaningful sample for which the detection and the status
  identification of the candidates was complete.

\end{itemize}

\subsection{Survey detection probability}

To correct for the projection effect from the observations, we then
ran a set of Monte-Carlo simulations using an optimized version of the
MESS code (Bonavita et al. 2012). For the \textit{full-stat} sample,
the code generates a uniform grid of mass and semi-major axis in the
interval [1, 75]~M$_{\rm{Jup}}$ and [1, 2000]~AU with a sampling of
0.5~M$_{\rm{Jup}}$ and 1~AU between 1 and 1000~AU, and 2~AU between
1000 and 2000~AU. For the \textit{complete-stat} sample, the uniform
grid is generated in the semi-major axis range between [1, 300]~AU
with a sampling of 1~AU. For each point in the grids, 100 orbits were
generated, randomly oriented in space from uniform distributions in
sin(i), $\omega$, $\Omega$, $e \le 0.8$, and $T_p$. The on-sky
projected position (separation and position angle) at the time of the
observation is then computed for each orbit and compared to our
\textbf{$5\sigma$} 2D-detection maps to determine the individual
detection probability ($p_j$) of planets around each star. The average
of all individual detection limits gives us the typical mean detection
probability ($<p_j>$) of the NaCo-LP to the planet and BD companion
population.  The results for the \textit{full-stat} and
\textit{complete-stat} samples are shown in Fig.~\ref{fig:detprob} and
~\ref{fig:upplim} \textit{Top}) respectively. The detection
probabilities in both cases do not significantly differ at less than
300~AU. Most companions more massive than 20~M$_{\rm{Jup}}$ with
semi-major axis between 70 and 200~AU should have been detected during
our survey. We are 50\% sensitive to massive ($\ge10$~M$_{\rm{Jup}}$)
planets and brown dwarfs with semi-major axis between 60 and
400~AU. Finally, the detection of giant planets as light as
5~M$_{\rm{Jup}}$ between 50-800~AU is only possible for 10\% of the
stars observed. The relatively small number of very young
  stars (see Fig.~\ref{fig:sample_prop}) is responsible for this
  limited sensitivity to light giant planets.

\subsection{Giant planet occurrence at wide orbits}

\begin{figure*}[t]
\includegraphics[width=9cm]{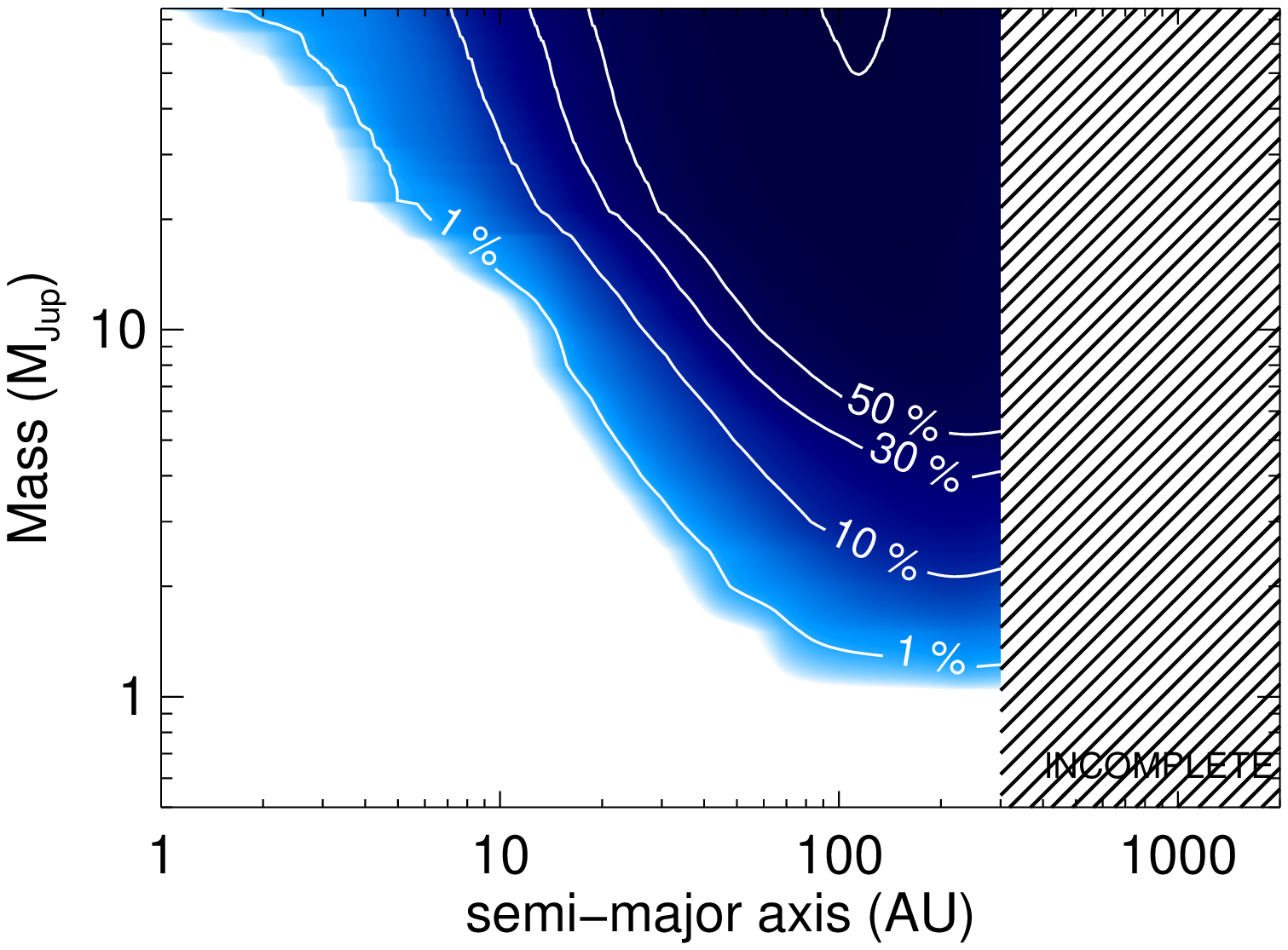}
\includegraphics[width=9cm]{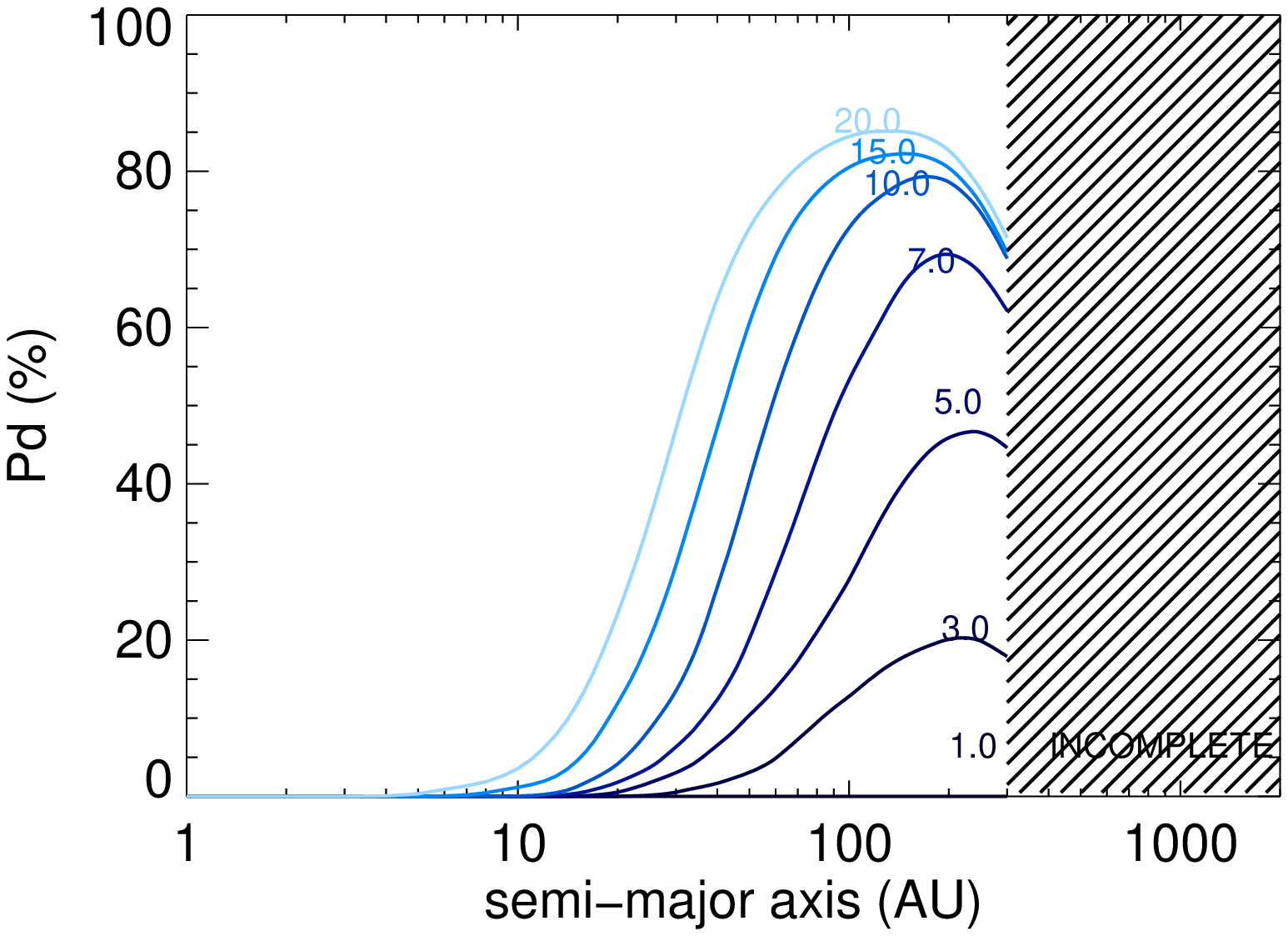}\\
\includegraphics[width=9cm]{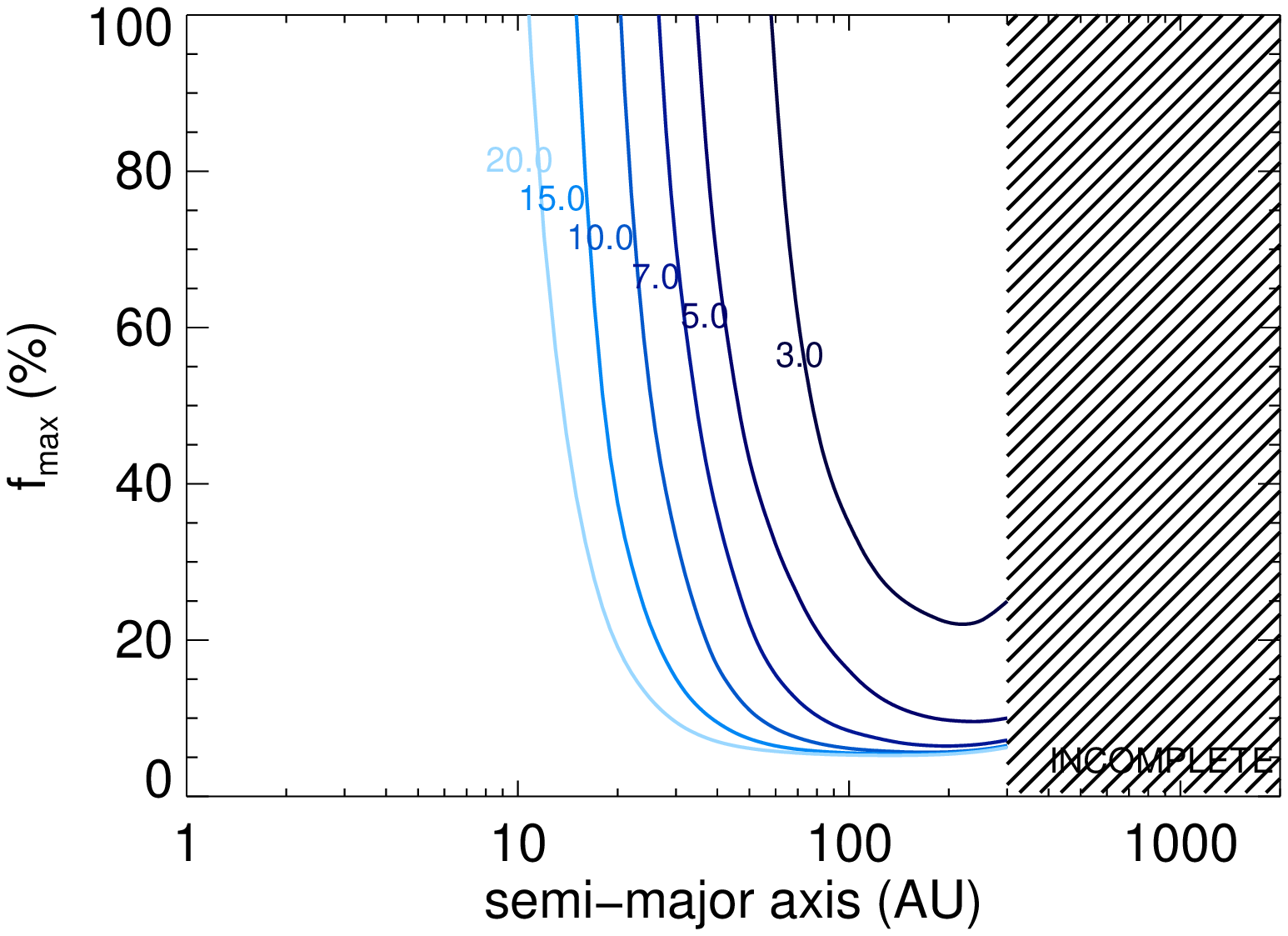}
\includegraphics[width=9cm]{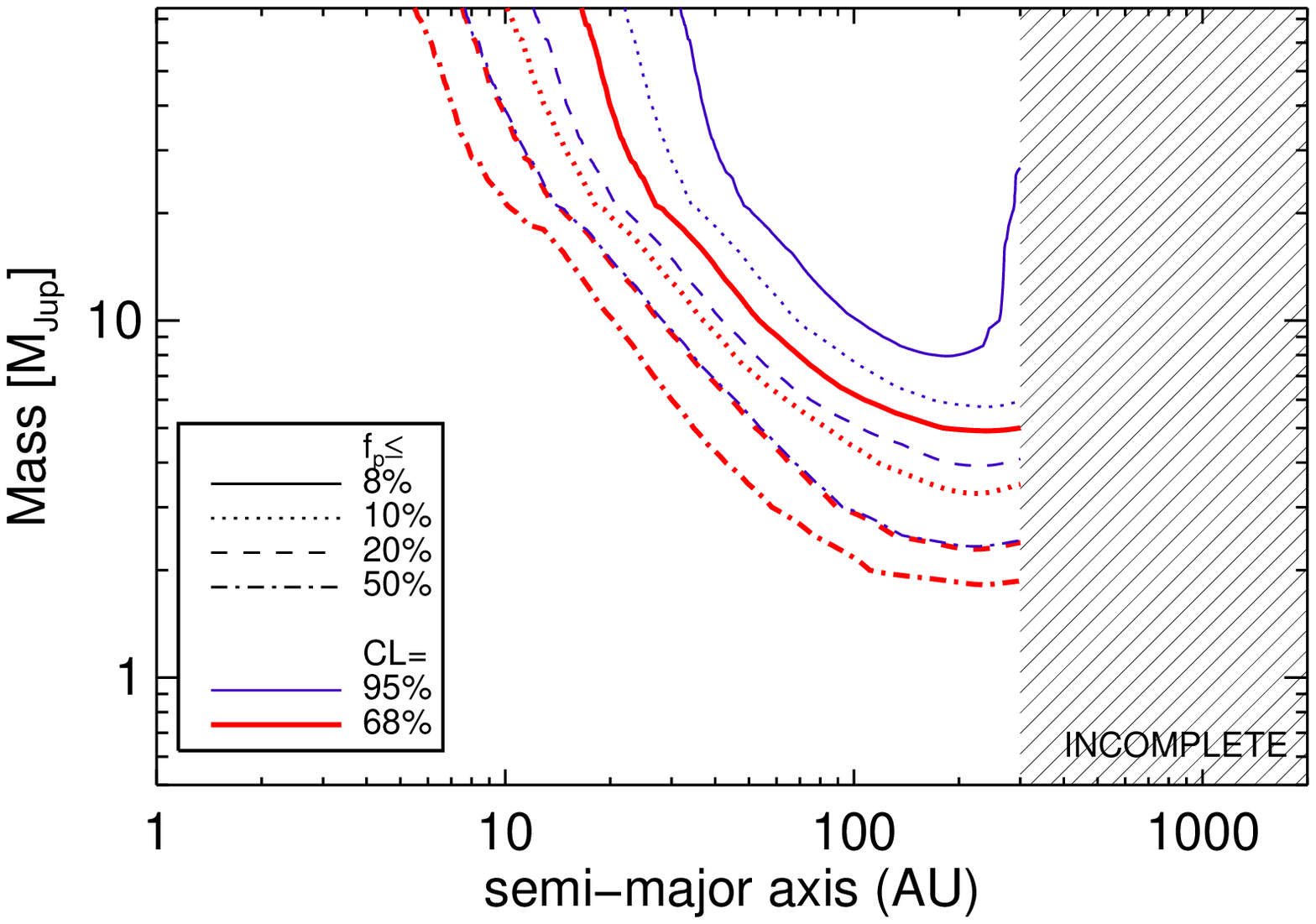}
\begin{centering}

\caption{Results for the \textit{complete-stat}
  sample. \textit{Top-Left:} NaCo-LP mean detection probability map
  ($<p_j>$) as a function of the mass and semi-major
  axis. \textit{Top-Right:} Mean probability curves for different
  masses (1, 3, 5, 7, 10, 15 and 20~M$_{\rm{Jup}}$) as a function of
  the semi-major axis. \textit{Bottom-Left:} Giant planet and brown
  dwarf occurrence upper limit ($f_{\rm{max}}$), considering a $95\%$
  confidence level, for different masses (3, 5, 7, 10, 15 and
  20~M$_{\rm{Jup}}$) as a function of the semi-major axis considering
  the null-detection result and an uniform distribution of planets and
  brown dwarfs in terms of masses and semi-major
  axis. \textit{Bottom-Right:} Same occurrence upper limit
  ($f_{\rm{max}}$) expressed this time in a mass versus semi-major
  axis diagramme for a 68\% and 95\% confidence level (following
  Biller et al. 2007; Nielsen et al. 2008 representation).}

\label{fig:upplim}
\end{centering}
\end{figure*}

To derive the occurrence of giant planets and brown dwarfs in our
survey, we only considered the \textit{complete-stat} sample with a
complete census of the candidates status within 300~AU. As no
planetary mass or brown dwarf companions were detected, we considered
here a null-detection result. We then used the mean detection
probability ($<p_j>$) to derive the giant planet and brown dwarf
occurrence upper limit ($f_{\rm{max}}$) compatible with the survey
detection limits. The probability of planet detection for a survey of
$N$ stars is described by a binomial distribution, given a success
probability $fp_j$ with $f$ the fraction of stars with planets. $p_j$
is the individual detection probability of detecting a planet if
present around the star $j$ and computed previously.  Assuming that
the number of expected detected planets is small compared to the
number of stars observed, the binomial distribution can be
approximated by a Poisson distribution to derive a simple analytical
solution for the exoplanet fraction upper limit ($f_{\rm{max}}$).  The
formalism is described by Carson et al. (2006) and Lafreni\`ere et
al. (2007). The result is shown in Fig.~\ref{fig:upplim}
(\textit{Bottom-Left} and \textit{Bottom-Right}). For this
\textit{complete-stat} sample, we constrain the occurrence of
exoplanets more massive than 5~M$_{\rm{Jup}}$ to typically less than
$15\%$ between 100 and 300~AU, and less than $10\%$ between 50 and
300~AU for exoplanets more massive than 10~M$_{\rm{Jup}}$ considering
a uniform input distribution and with a confidence level of
95\%. These values are consistent with current estimations from
various studies with comparable sensitivities around young, solar-type
stars \textbf{($f_{\rm{max}} \le 9.7\%$ for $[0.5,13]$~M$_{\rm{Jup}}$
  planet between $[50-250]$~AU by Lafreni\`ere et al. 2007;
  $f_{\rm{max}} \le 10\%$ for $[1,13]$~M$_{\rm{Jup}}$ planet between
  $[40-150]$~AU Chauvin et al. 2010; $f_{\rm{max}} \le 6\%$ for
  $[1,20]$~M$_{\rm{Jup}}$ planet between $[10-150]$~AU by Biller et
  al. 2013)}.

A more complete analysis combining the results of the NaCo-LP with
archive data for a total of $\sim210$ stars already observed in direct
imaging, will be presented in the related papers by Vigan et
al. (2014, in prep) and Reggiani et al. (2014, in prep). These
analysis will provide significant and relevant statistical constraints
on the population of planets and brown dwarfs around young, nearby
solar-type (FGK) stars (single or members of wide binaries) and enable
tests of planet and brown dwarf formation models.

\section{Conclusion}

In the context of the scientific preparation of the VLT/SPHERE
guaranteed time, we have conducted a survey of 86 young, nearby and
mostly solar-type stars using NaCo at VLT between 2009 and 2013. Our
main goals were to detect new giant planets and brown dwarf companions
and to initiate a relevant statistical study of their occurrence at
wide ($10-2000$~AU) orbit. NaCo was used in pupil-stabilized mode to
perform angular differential imaging at H-band. It enables us to reach
contrast performances as small as $10^{-6}$ at $1.5~\!''$. Over the 86
stars observed, the survey led to:

\begin{itemize}

\item the discovery of 11 new close binaries that we characterized in
  terms of relative photometry and astrometry.

\item the detection of more than 700 companion candidates, 90\% of
  them being located in six crowded fields. Among 76 stars observed in
  deep ADI, 33 systems have no point-source detected in their vicinity
  and 43 systems have at least one companion candidate
  detected. Repeated observations at several epochs enabled us to
  analyze the candidate status, completely or partially, around 29
  stars. Planetary mass candidates with proper follow-up were all
  identified as background sources. Additional follow-up observations
  will still be necessary to fully complete the status identification
  of all candidates detected in the survey owing to the variability of
  the detection performances from one run to another.  It shows that
  more than two epochs is generally necessary during a survey for a
  full exploration of the companions content.

\item the discovery of a unique comoving companion to the star
  HD\,8049. This result has been published by Zurlo et al. (2013) and
  revealed that the companion was actually a white dwarf with
  temperature $T_{\rm{eff}}=18800\pm2100$~K and mass M$_{\rm{WD}} =
  0.56\pm0.08~$M$_{\odot}$. 

\item new high-contrast images of the Moth debris-disk at
  HD\,61005. The NaCo $H$-band image remarkably resolves the disk
  component as a distinct narrow ring offset from the star by at least
  $2.75\pm0.85$~AU indicating a possibly dynamical perturbation by a
  planetary companion. This study was published by Buenzli et
  al. (2010).

\item finally, a preliminary statistical analysis of the survey
  detection probabitlities around the sample of 63 young, single and
  mostly solar-type (FGK) stars observed in angular differential
  imaging, i.e with detection performances enabling the search for
  planets and brown dwarfs in the stellar environment. Most companions
  more massive than 20~M$_{\rm{Jup}}$ with semi-major axis between 70
  and 200~AU should have been detected during our survey. We are 50\%
  sensitive to massive ($\ge10$~M$_{\rm{Jup}}$) planets and brown
  dwarfs with semi-major axis between 60 and 400~AU. Finally, the
  detection of giant planets as light as 5~M$_{\rm{Jup}}$ between
  50-800~AU is only possible for 10\% of the stars observed. We have
  then defined a more complete sample of 51 stars restreined to all
  systems for which the candidate status identification was complete
  up-to 300~AU, including cases with no companion candidates detected
  or with companion candidates properly and completely
  identified. Based on this complete sample average detection
  probability, a non-detection result and considering a uniform
  distribution of giant planets and brown dwarf companions in terms of
  semi-major axis and mass, we derive a typical upper limit for the
  occurrence of exoplanets more massive than 5~M$_{\rm{Jup}}$ of
  $15\%$ between 100 and 300~AU, and $10\%$ between 50 and 300~AU for
  EPs more massive than 10~M$_{\rm{Jup}}$ with a confidence level of
  95\%.

\end{itemize}

Combined with compiled archived data, the results of this survey offer
a unique sample of $\sim210$ young, solar-type stars observed in deep
imaging to constrain the presence of giant planets and brown dwarfs in
their close environment. A more complete statistical analysis will be
published in two linked articles by Vigan et al. (2014, in prep) and
Reggianni et al. (2014, in prep) to test the relevance of various
analytical distributions to describe the giant planet and brown dwarf
companion population at wide orbits, but also to bring further
constraints on current theories of planetary formation. All final
products of this survey (images, detection limits and candidate
status) will be released in the \textit{DIVA} (\textit{Deep Imaging
  Virtual Archive}) database, together with the archive data used for
full statistical analysis. We encourage the community to support this
effort by sharing the final products (reduced images, detection limits
and candidate relative astrometry, photometry and status) of their
published surveys to optimally prepare the future of planet imaging
searches coming with the new generation of planet imagers like
LMIRCam, MagAO, SPHERE, GPI, SCExAO and in a longer term JWST (Clampin
2010), TMT-PFI (Simard et al. 2010) and EELT-MIR and EELT-PCS (Brandl
et al. 2010; Kasper et al. 2010).

%
\bibliographystyle{aa}

\begin{acknowledgements}

We greatly thank the staff of ESO-VLT for their support at the
telescope. This publication has made use of the SIMBAD and VizieR
database operated at CDS, Strasbourg, France.  Finally, we acknowledge
supports from: 1) the French National Research Agency (ANR) through
project grant ANR10-BLANC0504-01 and the {\sl Programmes Nationaux de
  Plan\'etologie et de Physique Stellaire} (PNP \&\ PNPS), in France
for G.C., A.V, P.D., J-L. B., A-M. L. and D.M., 2) INAF through the
PRIN-INAF 2010 Planetary Systems at Young Ages project grant for S.D.,
D.M., M.B. and R.G. and 3) the U.S. National Science Foundation under
Award No. 1009203 for J.C.

\end{acknowledgements}


\begin{table*}[t]

\caption{Companion candidates characterization and identification (for
  multi-epoch observations). Target nameand observing date are given,
  as well as the different sources identified with their relative
  position and relative flux, and their identification status based on
  follow-up observations. Sources are indicated as: stationary
  background contaminants (B; based on a comoving companion
  probability P$_{\rm{comoving}, \chi^2}<1$\% and with a relative
  motion compatible with a background source); confirmed comoving
  companions (C; based on a stationary background contaminant
  probability P$_{\rm{BKG}, \chi^2}<1$\% and a relative motion
  compatible with a comoving companion); and undefined (U; when
  observed at only one epoch or when not satisfying the first two
  classifications).}

\begin{center}
\small
\begin{tabular}{llllllll}     
\noalign{\smallskip}
\noalign{\smallskip}\hline
Name-1                        &     UT-Date  & Candidate    &    Sep      &     PA    &  $\Delta H$   &    Status     &    Comments     \\
                               &              &             & (mas)       & (deg)     & (mag)         &               &                 \\
\noalign{\smallskip}\hline  \noalign{\smallskip}
 TYC\,5839-0596-1        & 2009-11-24 &       none & & & &    & SB2  \\
 TYC\,0603-0461-1        & 2009-11-24 &       none & & & &    & New binary (see Table~6) \\
        HIP3924          & 2009-11-22 &       none & & & &    & SB2 \\
        HIP6177          & 2010-07-31 &    cc-1    & $1566\pm6$  & $118.4\pm0.2$ & $7.1\pm0.1$ & &  \\
                         & 2011-07-28 &    cc-1    & $1565\pm10$ & $118.0\pm0.4$ & & C  & White dwarf companion\\  
        HIP8038          & 2010-07-31 &       none & & & &    & New binary (see Table~6) \\
       HIP10602          & 2009-11-24 &       none & & & &    & A few exposures\\
                         & 2010-07-30 &       none & & & &    & \\
       HIP11360          & 2009-11-23 &       none & & & &    & \\
 TYC\,8484-1507-1        & 2010-07-31 &       none & & & &    & Known ($\sim8.6\!''$) binary  \\
                         &            &            & & & &    & resolved by 2MASS \\
       HIP12394          & 2009-11-22 &       none & & &    &  \\
         HIP13008        & 2011-09-29 &    cc-1     & $   1710 \pm 7   $ & $347.3 \pm 0.2 $ & $    6.9 \pm 0.0 $ &   U &      \\ 
       HIP14684          & 2010-07-30 &    cc-1 & $   5454 \pm 13  $ & $150.6 \pm 0.1 $ & $   10.7 \pm 0.1 $ &   B &      \\ 
                         & 2011-12-23 &    cc-1 & $   5274 \pm 6   $ & $150.3 \pm 0.1 $ & $   11.1 \pm 0.1 $ &   B &      \\ 
 TYC\,8060-1673-1        & 2009-11-23 &        none & & &    &  \\
       HIP19775          & 2009-11-22 &        none & & &    &  \\
       HIP23316          & 2009-11-23 &        none & & &    &   \\
        HD32981          & 2009-11-24 &        none & & &    &  \\
      BD-09-1108         & 2009-11-22 &        none & & &    &  \\
       HIP25434          & 2010-02-17 &    cc-1 & $   4944 \pm 11  $ & $154.5 \pm 0.1 $ & $   11.4 \pm 0.1 $ &   B &      \\ 
                         & 2010-12-05 &    cc-1 & $   4937 \pm 5   $ & $154.5 \pm 0.1 $ & $   12.0 \pm 0.1 $ &   B &      \\ 
                         & 2012-11-21 &    cc-1 & $   4947 \pm 7   $ & $155.1 \pm 0.1 $ & $   11.1 \pm 0.1 $ &   B &      \\ 
 TYC\,9162-0698-1        & 2010-02-19 &$26$     & & & &    & Electronic table  \\
                         & 2011-01-24 &$26+33$  & & & & B+U   & Electronic table \\
  TYC\,5346-132-1        & 2009-11-23 &    cc-1 & $   6252 \pm 16  $ & $  1.8 \pm 0.2 $ & $    9.7 \pm 0.1 $ &   B &      \\ 
                         & 2010-02-16 &    cc-1 & $   6260 \pm 9   $ & $  1.7 \pm 0.1 $ & $    9.8 \pm 0.1 $ &   B &      \\ 
                         & 2009-11-23 &    cc-2 & $   6431 \pm 16  $ & $  0.2 \pm 0.2 $ & $    9.1 \pm 0.1 $ &   B &      \\ 
                         & 2010-02-16 &    cc-2 & $   6434 \pm 9   $ & $  0.0 \pm 0.1 $ & $    9.1 \pm 0.1 $ &   B &      \\ 
       HIP30261          & 2009-11-23 &        none & & &    &  \\
 TYC\,7617-0549-1        & 2009-11-21 &    cc-1 & $   1848 \pm 16  $ & $299.6 \pm 0.5 $ & $   12.6 \pm 0.1 $ &   B &      \\ 
                         & 2012-11-22 &    cc-1 & $   1861 \pm 8   $ & $298.6 \pm 0.3 $ & $   12.1 \pm 0.1 $ &   B &      \\ 
 TYC\,9181-0466-1        & 2010-02-19 &        none & & &    &  & New binary  (see Table~6) \\
       HIP32235          & 2010-02-18 &    cc-1 & $   5559 \pm 12  $ & $340.3 \pm 0.2 $ & $   11.8 \pm 0.1 $ &   B &      \\ 
                         & 2010-12-30 &    cc-1 & $   5508 \pm 6   $ & $339.8 \pm 0.1 $ & $   13.3 \pm 0.2 $ &   B &      \\ 
       HIP35564          & 2009-11-22 &    cc-1 & $   1865 \pm 20  $ & $304.2 \pm 0.6 $ & $   15.3 \pm 0.3 $ &   B & RV var     \\ 
                 & 2011-01-31 &    cc-1 & $   1852 \pm 8   $ & $299.3 \pm 0.2 $ & $   14.4 \pm 0.7 $ &   B &      \\ 
                 & 2009-11-22 &    cc-2 & $   3301 \pm 20  $ & $148.9 \pm 0.3 $ & $   12.4 \pm 0.1 $ &   B &      \\ 
                 & 2010-02-16 &    cc-2 & $   3365 \pm 13  $ & $148.7 \pm 0.2 $ & $   12.1 \pm 0.1 $ &   B &      \\ 
                 & 2011-01-31 &    cc-2 & $   3465 \pm 9   $ & $149.8 \pm 0.2 $ & $   11.9 \pm 0.1 $ &   B &      \\ 
                 & 2009-11-22 &    cc-3 & $   6722 \pm 20  $ & $  1.5 \pm 0.2 $ & $   14.4 \pm 0.3 $ &   B &      \\ 
                 & 2010-02-16 &    cc-3 & $   6660 \pm 12  $ & $  1.8 \pm 0.2 $ & $   14.0 \pm 0.4 $ &   B &      \\ 
                 & 2011-01-31 &    cc-3 & $   6581 \pm 8   $ & $  1.7 \pm 0.1 $ & $   13.7 \pm 0.4 $ &   B &      \\ 
 TYC\,8128-1946-1        & 2009-11-21 &    cc-1 & $   5521 \pm 17  $ & $178.2 \pm 0.2 $ & $   13.5 \pm 0.1 $ &   B &      \\ 
                 & 2011-01-20 &    cc-1 & $   5549 \pm 12  $ & $178.1 \pm 0.2 $ & $   13.7 \pm 0.2 $ &   B &      \\ 
                 & 2009-11-21 &    cc-2 & $   8211 \pm 17  $ & $  6.3 \pm 0.2 $ & $    9.4 \pm 0.1 $ &   B &      \\ 
                 & 2011-01-20 &    cc-2 & $   8190 \pm 12  $ & $  6.3 \pm 0.1 $ & $    9.9 \pm 0.1 $ &   B &      \\ 
       HIP36414          & 2010-02-17 &    cc-1 & $   8296 \pm 21  $ & $305.0 \pm 0.2 $ & $   12.8 \pm 0.7 $ &   B & SB, RV var     \\ 
                 & 2011-01-31 &    cc-1 & $   8241 \pm 16  $ & $304.5 \pm 0.1 $ & $   13.5 \pm 0.2 $ &   B &      \\ 
                 & 2010-02-17 &    cc-2 & $   7076 \pm 19  $ & $359.2 \pm 0.2 $ & $   13.6 \pm 0.2 $ &   U &      \\ 
       HIP36948          & 2010-02-16 &    cc-1 & $   3485 \pm 21  $ & $327.1 \pm 0.3 $ & $   14.1 \pm 0.2 $ &   B & The Moth$^a$ \\ 
                 & 2010-02-16 &    cc-2 & $   6272 \pm 22  $ & $315.5 \pm 0.2 $ & $   13.6 \pm 0.2 $ &   B &      \\ 
                 & 2010-02-16 &    cc-3 & $   7217 \pm 20  $ & $191.3 \pm 0.2 $ & $   13.3 \pm 0.4 $ &   B &      \\ 
                 & 2010-02-16 &    cc-4 & $   8116 \pm 20  $ & $171.1 \pm 0.2 $ & $   14.1 \pm 0.8 $ &   B &      \\ 
                 & 2010-02-16 &    cc-5 & $   8206 \pm 20  $ & $268.1 \pm 0.2 $ & $   10.1 \pm 0.1 $ &   B & \\ 
       HIP37563  & 2010-02-18 & none    &                    &                  &                    &       & \\
\noalign{\smallskip}\hline  \noalign{\smallskip}
\end{tabular}
\end{center}
\begin{list}{}{}
\item[\scriptsize{(a):}] \scriptsize{All background objects identified combining NaCo with HST data by Buenzli et al. (2010)}
\end{list}
\end{table*}

\begin{table*}[p]
\caption{Companion candidates characterization and identification (for multi-epoch
  observations). Table~7-cont.}
\begin{center}
\small
\begin{tabular}{llllllll}     
\noalign{\smallskip}
\noalign{\smallskip}\hline
Name-1           &     UT-Date  & Nb Cand.    &     Sep      &     PA    &  $\Delta H$   &    Status     &    Comments     \\
                 &              &                 & (mas)         & (deg)     & (mag)         &               &                 \\
\noalign{\smallskip}\hline  \noalign{\smallskip}
       HIP37923  & 2010-02-18 &    cc-1 & $   5439 \pm 10  $ & $261.2 \pm 0.2 $ & $   12.6 \pm 0.1 $ &   B &      \\ 
                 & 2011-01-01 &    cc-1 & $   5439 \pm 8   $ & $259.8 \pm 0.1 $ & $   12.7 \pm 0.1 $ &   B &      \\ 
                 & 2010-02-18 &    cc-2 & $   5834 \pm 12  $ & $ 55.5 \pm 0.1 $ & $   14.0 \pm 0.1 $ &   B &      \\ 
                 & 2011-01-01 &    cc-2 & $   5765 \pm 10  $ & $ 56.5 \pm 0.1 $ & $   14.2 \pm 0.2 $ &   B &      \\ 
                 & 2010-02-18 &    cc-3 & $   5997 \pm 12  $ & $209.9 \pm 0.1 $ & $   12.4 \pm 0.1 $ &   B &      \\ 
                 & 2011-01-01 &    cc-3 & $   6098 \pm 10  $ & $209.1 \pm 0.1 $ & $   12.7 \pm 0.1 $ &   B &      \\ 
                 & 2010-02-18 &    cc-4 & $   7070 \pm 12  $ & $ 28.2 \pm 0.1 $ & $   14.1 \pm 0.3 $ &   B &      \\ 
                 & 2011-01-01 &    cc-4 & $   6947 \pm 10  $ & $ 28.8 \pm 0.1 $ & $   14.4 \pm 0.2 $ &   B &      \\ 
                 & 2010-02-18 &    cc-5 & $   8076 \pm 13  $ & $ 65.1 \pm 0.1 $ & $    9.1 \pm 0.1 $ &   B &      \\ 
                 & 2011-01-01 &    cc-5 & $   8029 \pm 10  $ & $ 65.9 \pm 0.1 $ & $    9.6 \pm 0.1 $ &   B &      \\ 
                 & 2010-02-18 &    cc-6 & $   8677 \pm 15  $ & $ 43.1 \pm 0.1 $ & $   13.8 \pm 0.2 $ &   U &      \\ 
 TYC\,8927-3620-1 & 2010-02-19 &        none & & & &   &  New binary (see Table~6)\\
                  &            &             & & & &   &  Third component at $\sim4.8~\!''$ \\
                  &            &             & & & &   &  resolved by 2MASS \\
       HIP46634   & 2009-11-24 &        none & & & &   &  \\
       HIP47646   & 2010-02-18 &        none & & & &   &  \\
          TWA-21  & 2010-02-18 &    cc-1 & $   2353 \pm 11  $ & $ 30.9 \pm 0.3 $ & $   13.8 \pm 0.3 $ &   B &      \\ 
                 & 2011-01-13 &    cc-1 & $   2339 \pm 7   $ & $ 31.5 \pm 0.2 $ & $   14.2 \pm 0.4 $ &   B &      \\ 
                 & 2010-02-18 &    cc-2 & $   2508 \pm 11  $ & $342.2 \pm 0.3 $ & $   13.9 \pm 0.3 $ &   B &      \\ 
                 & 2011-01-13 &    cc-2 & $   2489 \pm 7   $ & $342.7 \pm 0.2 $ & $   14.2 \pm 0.3 $ &   B &      \\ 
                 & 2010-02-18 &    cc-3 & $   3152 \pm 11  $ & $ 67.6 \pm 0.2 $ & $   12.2 \pm 0.1 $ &   B &      \\ 
                 & 2011-01-13 &    cc-3 & $   3178 \pm 7   $ & $ 68.1 \pm 0.2 $ & $   12.8 \pm 0.1 $ &   B &      \\ 
                 & 2010-02-18 &    cc-4 & $   4968 \pm 11  $ & $ 94.0 \pm 0.2 $ & $   13.4 \pm 0.2 $ &   B &      \\ 
                 & 2011-01-13 &    cc-4 & $   5003 \pm 6   $ & $ 94.1 \pm 0.1 $ & $   14.0 \pm 0.3 $ &   B &      \\ 
                 & 2010-02-18 &    cc-5 & $   5231 \pm 12  $ & $241.6 \pm 0.2 $ & $    8.9 \pm 0.1 $ &   B &      \\ 
                 & 2011-01-13 &    cc-5 & $   5210 \pm 8   $ & $241.2 \pm 0.1 $ & $    9.4 \pm 0.1 $ &   B &      \\ 
                 & 2010-02-18 &    cc-6 & $   5355 \pm 12  $ & $ 71.1 \pm 0.2 $ & $   11.7 \pm 0.1 $ &   B &      \\ 
                 & 2011-01-13 &    cc-6 & $   5387 \pm 7   $ & $ 71.3 \pm 0.1 $ & $   12.4 \pm 0.1 $ &   B &      \\ 
                 & 2010-02-18 &    cc-7 & $   5602 \pm 12  $ & $ 24.6 \pm 0.1 $ & $    9.1 \pm 0.1 $ &   B &      \\ 
                 & 2011-01-13 &    cc-7 & $   5599 \pm 8   $ & $ 25.0 \pm 0.1 $ & $    9.7 \pm 0.1 $ &   B &      \\ 
                 & 2010-02-18 &    cc-8 & $   5712 \pm 12  $ & $ 27.8 \pm 0.1 $ & $   12.7 \pm 0.1 $ &   B &      \\ 
                 & 2011-01-13 &    cc-8 & $   5717 \pm 9   $ & $ 28.0 \pm 0.1 $ & $   13.2 \pm 0.1 $ &   B &      \\ 
                 & 2010-02-18 &    cc-9 & $   5801 \pm 12  $ & $245.6 \pm 0.1 $ & $   10.2 \pm 0.1 $ &   B &      \\ 
                 & 2011-01-13 &    cc-9 & $   5776 \pm 8   $ & $245.3 \pm 0.1 $ & $   10.7 \pm 0.1 $ &   B &      \\ 
                 & 2010-02-18 &   cc-10 & $   5833 \pm 11  $ & $102.6 \pm 0.2 $ & $   13.4 \pm 0.2 $ &   B &      \\ 
                 & 2011-01-13 &   cc-10 & $   5879 \pm 7   $ & $102.6 \pm 0.1 $ & $   14.0 \pm 0.2 $ &   B &      \\ 
                 & 2010-02-18 &   cc-11 & $   6097 \pm 12  $ & $107.7 \pm 0.1 $ & $   12.6 \pm 0.1 $ &   B &      \\ 
                 & 2011-01-13 &   cc-11 & $   6138 \pm 7   $ & $107.7 \pm 0.1 $ & $   13.3 \pm 0.1 $ &   B &      \\ 
                 & 2010-02-18 &   cc-12 & $   6951 \pm 13  $ & $147.5 \pm 0.1 $ & $   13.6 \pm 0.2 $ &   B &      \\ 
                 & 2011-01-13 &   cc-12 & $   6990 \pm 10  $ & $147.4 \pm 0.1 $ & $   14.4 \pm 0.4 $ &   B &      \\ 
                 & 2010-02-18 &   cc-13 & $   3369 \pm 11  $ & $165.3 \pm 0.2 $ & $   14.3 \pm 0.3 $ &   U &      \\ 
                 & 2010-02-18 &   cc-14 & $   5948 \pm 13  $ & $228.5 \pm 0.1 $ & $   15.0 \pm 0.4 $ &   U &      \\ 
                 & 2011-01-13 &   cc-15 & $   7049 \pm 7   $ & $  9.4 \pm 0.1 $ & $   12.9 \pm 0.1 $ &   U &      \\
 TYC\,7188-0575-1 & 2010-02-16 &    cc-1 & $   4238 \pm 15  $ & $296.5 \pm 0.2 $ & $   15.2 \pm 0.4 $ &   B & SB2     \\ 
                 & 2011-01-27 &    cc-1 & $   4148 \pm 12  $ & $297.5 \pm 0.2 $ & $   14.8 \pm 0.4 $ &   B &      \\ 
                 & 2010-02-16 &    cc-2 & $   4741 \pm 15  $ & $ 61.7 \pm 0.2 $ & $   11.6 \pm 0.1 $ &   B &      \\ 
                 & 2011-01-27 &    cc-2 & $   4843 \pm 12  $ & $ 62.2 \pm 0.2 $ & $   11.4 \pm 0.1 $ &   B &      \\ 
                 & 2010-02-16 &    cc-3 & $   5329 \pm 15  $ & $186.0 \pm 0.2 $ & $   11.9 \pm 0.1 $ &   B &      \\ 
                 & 2011-01-27 &    cc-3 & $   5303 \pm 11  $ & $185.1 \pm 0.2 $ & $   11.8 \pm 0.1 $ &   B &      \\ 
                 & 2010-02-16 &    cc-4 & $   7391 \pm 15  $ & $279.5 \pm 0.2 $ & $   13.2 \pm 0.2 $ &   B &      \\ 
                 & 2011-01-27 &    cc-4 & $   7282 \pm 11  $ & $279.8 \pm 0.1 $ & $   13.1 \pm 0.2 $ &   B &      \\ 
                 & 2010-02-16 &    cc-5 & $   8020 \pm 15  $ & $  5.1 \pm 0.2 $ & $   11.6 \pm 0.1 $ &   B &      \\ 
                 & 2011-01-27 &    cc-5 & $   8058 \pm 11  $ & $  5.8 \pm 0.1 $ & $   11.7 \pm 0.1 $ &   B &      \\ 
 TYC\,6069-1214-1 & 2010-02-17 &        none & & &    &  \\
 TYC\,7722-0207-1 & 2010-02-17 &    cc-1 & $   3981 \pm 11  $ & $ 33.5 \pm 0.2 $ & $   11.7 \pm 0.1 $ &   B &      \\ 
                 & 2011-01-29 &    cc-1 & $   3978 \pm 6   $ & $ 35.0 \pm 0.1 $ & $   11.5 \pm 0.1 $ &   B &      \\ 
                 & 2010-02-17 &    cc-2 & $   4369 \pm 12  $ & $228.1 \pm 0.2 $ & $    7.3 \pm 0.1 $ &   B &      \\ 
                 & 2011-01-29 &    cc-2 & $   4355 \pm 7   $ & $226.9 \pm 0.1 $ & $    7.1 \pm 0.1 $ &   B &      \\ 
                 & 2010-02-17 &    cc-3 & $   8516 \pm 11  $ & $105.2 \pm 0.1 $ & $    9.4 \pm 0.1 $ &   B &      \\ 
                 & 2011-01-29 &    cc-3 & $   8602 \pm 6   $ & $105.4 \pm 0.1 $ & $    9.1 \pm 0.1 $ &   B &      \\ 
                 & 2010-02-17 &    cc-4 & $   1742 \pm 10  $ & $329.8 \pm 0.3 $ & $   13.9 \pm 0.5 $ &   U &      \\ 
                 & 2011-01-29 &    cc-5 & $   7958 \pm 11  $ & $317.3 \pm 0.1 $ & $   10.1 \pm 0.1 $ &   U &      \\ 
                 & 2011-01-29 &    cc-6 & $   7988 \pm 11  $ & $ 42.2 \pm 0.1 $ & $   12.4 \pm 0.1 $ &   U &      \\ 
 \noalign{\smallskip}\hline                  \noalign{\smallskip}
\end{tabular}
\end{center}
\end{table*}

\begin{table*}[p]
\caption{Companion candidates characterization and identification (for multi-epoch
  observations). Table~8-cont.}
\begin{center}
\small
\begin{tabular}{llllllll}     
\noalign{\smallskip}
\noalign{\smallskip}\hline
Name-1           &     UT-Date  & Nb Cand.    &     Sep      &     PA    &  $\Delta H$   &    Status     &    Comments     \\
                 &              &                 & (mas)         & (deg)     & (mag)         &               &                 \\
\noalign{\smallskip}\hline  \noalign{\smallskip}
 TYC\,7743-1091-1 & 2010-02-19 &        none & & &    &  & \\
       HIP58240   & 2010-02-16 &    cc-1 & $   5761 \pm 21  $ & $179.6 \pm 0.2 $ & $   13.0 \pm 0.2 $ &   B &      \\ 
                 & 2011-01-29 &    cc-1 & $   5770 \pm 5   $ & $178.1 \pm 0.1 $ & $   12.9 \pm 0.1 $ &   B &      \\ 
 TYC\,9231-1566-1 & 2010-02-19   & none           & & & &    &  New binary (see Table~6)  \\
 TYC\,8979-1683-1 & 2010-02-18   & $54(+16)$       & & & &    & Electronic table  \\
                  & 2011-05-11   & $54$       & & & & B(+U) & Electronic table \\
 TYC\,8989-0583-1 & 2010-02-18   &           none & & & &    &  New binary (see Table~6) \\
                  & 2010-06-16   &           none & & & &    &  \\
 TYC\,9245-0617-1 & 2010-02-18 &    cc-1 & $   3771 \pm 10  $ & $ 32.2 \pm 0.2 $ & $    9.5 \pm 0.1 $ &   B &      \\ 
                 & 2013-02-11 &    cc-1 & $   3860 \pm 5   $ & $ 33.2 \pm 0.1 $ & $    9.1 \pm 0.1 $ &   B &      \\ 
                 & 2010-02-18 &    cc-2 & $   3942 \pm 10  $ & $119.4 \pm 0.2 $ & $    5.9 \pm 0.1 $ &   B &      \\ 
                 & 2011-04-04 &    cc-2 & $   3974 \pm 10  $ & $118.9 \pm 0.2 $ & $    6.0 \pm 0.1 $ &   B &      \\ 
                 & 2013-02-11 &    cc-2 & $   4015 \pm 5   $ & $118.4 \pm 0.1 $ & $    5.8 \pm 0.1 $ &   B &      \\ 
                 & 2010-02-18 &    cc-3 & $   6544 \pm 9   $ & $183.4 \pm 0.1 $ & $   10.9 \pm 0.1 $ &   B &      \\ 
                 & 2013-02-11 &    cc-3 & $   6517 \pm 3   $ & $182.4 \pm 0.1 $ & $   10.9 \pm 0.3 $ &   B &      \\ 
                 & 2010-02-18 &    cc-4 & $   7448 \pm 11  $ & $241.0 \pm 0.1 $ & $   10.2 \pm 0.1 $ &   B &      \\ 
                 & 2013-02-11 &    cc-4 & $   7342 \pm 8   $ & $241.0 \pm 0.1 $ & $   10.2 \pm 0.2 $ &   B &      \\ 
                 & 2010-02-18 &    cc-5 & $   4590 \pm 10  $ & $306.1 \pm 0.1 $ & $   12.8 \pm 0.1 $ &   U &      \\ 
                 & 2010-02-18 &    cc-6 & $   5603 \pm 9   $ & $263.3 \pm 0.1 $ & $   14.5 \pm 0.3 $ &   U &      \\ 
                 & 2010-02-18 &    cc-7 & $   5887 \pm 9   $ & $179.7 \pm 0.1 $ & $   12.2 \pm 0.1 $ &   U &      \\ 
                 & 2010-02-18 &    cc-8 & $   6149 \pm 10  $ & $251.4 \pm 0.1 $ & $   13.3 \pm 0.2 $ &   U &      \\ 
                 & 2010-02-18 &    cc-9 & $   7432 \pm 12  $ & $146.0 \pm 0.1 $ & $   11.9 \pm 0.1 $ &   U &      \\ 
                 & 2010-02-18 &   cc-10 & $   8529 \pm 10  $ & $ 73.1 \pm 0.1 $ & $   12.9 \pm 0.2 $ &   U &      \\ 
       HIP63862 & 2010-02-18 &    cc-1 & $   4231 \pm 16  $ & $ 28.0 \pm 0.2 $ & $   12.9 \pm 0.1 $ &   B &      \\ 
                 & 2011-07-02 &    cc-1 & $   4315 \pm 5   $ & $ 30.7 \pm 0.1 $ & $   13.1 \pm 0.1 $ &   B &      \\ 
                 & 2010-02-18 &    cc-2 & $   5536 \pm 17  $ & $206.7 \pm 0.2 $ & $   12.8 \pm 0.1 $ &   B &      \\ 
                 & 2011-07-02 &    cc-2 & $   5478 \pm 5   $ & $204.8 \pm 0.1 $ & $   13.3 \pm 0.1 $ &   B &      \\ 
                 & 2011-07-02 &    cc-3 & $   7186 \pm 5   $ & $160.1 \pm 0.1 $ & $   14.1 \pm 0.2 $ &   U &      \\ 
                 & 2011-07-02 &    cc-4 & $   8169 \pm 10  $ & $142.8 \pm 0.1 $ & $   12.7 \pm 0.1 $ &   U &      \\  
 TYC\,7796-2110-1 & 2010-02-16 &    cc-1 & $   3264 \pm 15  $ & $127.0 \pm 0.3 $ & $   11.3 \pm 0.1 $ &   B &      \\ 
                 & 2011-05-11 &    cc-1 & $   3306 \pm 4   $ & $126.3 \pm 0.1 $ & $   11.1 \pm 0.3 $ &   B &      \\ 
                 & 2013-03-22 &    cc-1 & $   3314 \pm 9   $ & $125.2 \pm 0.2 $ & $   11.2 \pm 0.1 $ &   B &      \\ 
                 & 2010-02-16 &    cc-2 & $   3990 \pm 16  $ & $141.6 \pm 0.2 $ & $    8.7 \pm 0.1 $ &   B &      \\ 
                 & 2011-05-11 &    cc-2 & $   4018 \pm 5   $ & $141.1 \pm 0.1 $ & $    8.8 \pm 0.1 $ &   B &      \\ 
                 & 2013-03-22 &    cc-2 & $   4034 \pm 9   $ & $140.3 \pm 0.1 $ & $    8.7 \pm 0.1 $ &   B &      \\ 
                 & 2010-02-16 &    cc-3 & $   4111 \pm 15  $ & $109.7 \pm 0.2 $ & $   13.5 \pm 0.2 $ &   B &      \\ 
                 & 2013-03-22 &    cc-3 & $   4190 \pm 8   $ & $108.5 \pm 0.1 $ & $   13.6 \pm 0.3 $ &   B &      \\ 
                 & 2010-02-16 &    cc-4 & $   5231 \pm 16  $ & $151.6 \pm 0.2 $ & $   13.2 \pm 0.1 $ &   B &      \\ 
                 & 2013-03-22 &    cc-4 & $   5194 \pm 10  $ & $150.7 \pm 0.1 $ & $   13.3 \pm 0.2 $ &   B &      \\ 
                 & 2010-02-16 &    cc-5 & $   6324 \pm 17  $ & $214.1 \pm 0.2 $ & $   12.5 \pm 0.1 $ &   B &      \\ 
                 & 2013-03-22 &    cc-5 & $   6211 \pm 11  $ & $214.0 \pm 0.1 $ & $   12.4 \pm 0.1 $ &   B &      \\ 
                 & 2010-02-16 &    cc-6 & $   7111 \pm 16  $ & $ 62.6 \pm 0.2 $ & $   10.8 \pm 0.1 $ &   B &      \\ 
                 & 2011-05-11 &    cc-6 & $   7145 \pm 7   $ & $ 62.8 \pm 0.1 $ & $   11.1 \pm 0.3 $ &   B &      \\ 
                 & 2013-03-22 &    cc-6 & $   7240 \pm 10  $ & $ 62.8 \pm 0.1 $ & $   10.8 \pm 0.1 $ &   B &      \\ 
                 & 2010-02-16 &    cc-7 & $   8743 \pm 18  $ & $ 48.5 \pm 0.1 $ & $   14.0 \pm 0.4 $ &   U &      \\ 
 TYC\,9010-1272-1 & 2010-02-18   &           none & & & &    &  New binary (see Table~6) \\
       HIP70351   & 2010-02-17 &    cc-1 & $   2971 \pm 10  $ & $258.4 \pm 0.2 $ & $   14.7 \pm 0.4 $ &   U &      \\ 
                 & 2010-02-17 &    cc-2 & $   4664 \pm 11  $ & $288.7 \pm 0.2 $ & $   11.4 \pm 0.1 $ &   U &      \\ 
                 & 2010-02-17 &    cc-3 & $   4973 \pm 10  $ & $  4.0 \pm 0.2 $ & $   13.9 \pm 0.2 $ &   U &      \\ 
                 & 2010-02-17 &    cc-4 & $   6572 \pm 10  $ & $262.2 \pm 0.1 $ & $   13.5 \pm 0.1 $ &   U &      \\ 
                 & 2010-02-17 &    cc-5 & $   6615 \pm 13  $ & $235.8 \pm 0.1 $ & $   13.0 \pm 0.1 $ &   U &      \\ 
                 & 2010-02-17 &    cc-6 & $   6820 \pm 13  $ & $145.7 \pm 0.1 $ & $   11.0 \pm 0.1 $ &   U &      \\ 
                 & 2010-02-17 &    cc-7 & $   7266 \pm 12  $ & $118.2 \pm 0.1 $ & $   12.9 \pm 0.1 $ &   U &      \\ 
       HIP71908   & 2010-02-16 &    cc-1 & $   6404 \pm 16  $ & $ 28.4 \pm 0.2 $ & $   16.3 \pm 0.4 $ &   U &      \\ 
       HIP71933   & 2010-06-15 &    cc-1 & $   4934 \pm 9   $ & $ 12.6 \pm 0.1 $ & $   10.0 \pm 0.1 $ &   B &      \\ 
                 & 2011-04-06 &    cc-1 & $   4952 \pm 2   $ & $ 12.8 \pm 0.1 $ & $    9.4 \pm 0.1 $ &   B &      \\ 
                 & 2013-02-23 &    cc-1 & $   5031 \pm 3   $ & $ 13.1 \pm 0.1 $ & $   10.0 \pm 0.1 $ &   B &      \\ 
                 & 2010-06-15 &    cc-2 & $   5864 \pm 11  $ & $ 54.1 \pm 0.1 $ & $   11.3 \pm 0.1 $ &   B &      \\ 
                 & 2013-02-23 &    cc-2 & $   5958 \pm 7   $ & $ 54.0 \pm 0.1 $ & $   11.3 \pm 0.1 $ &   B &      \\ 
 \noalign{\smallskip}\hline                  \noalign{\smallskip}
\end{tabular}
\end{center}
\end{table*}

\begin{table*}[p]
\caption{Companion candidates characterization and identification (for multi-epoch
  observations). Table~9-cont.}
\begin{center}
\small
\begin{tabular}{llllllll}     
\noalign{\smallskip}
\noalign{\smallskip}\hline
Name-1           &     UT-Date  & Nb Cand.    &     Sep      &     PA    &  $\Delta H$   &    Status     &    Comments     \\
                 &              &                 & (mas)         & (deg)     & (mag)         &               &                 \\
\noalign{\smallskip}\hline  \noalign{\smallskip}
       HIP71933  & 2010-06-15 &    cc-3 & $   7987 \pm 9   $ & $ 12.5 \pm 0.1 $ & $   10.4 \pm 0.1 $ &   B &      \\ 
                 & 2011-04-06 &    cc-3 & $   8006 \pm 3   $ & $ 12.8 \pm 0.1 $ & $    9.8 \pm 0.2 $ &   B &      \\ 
                 & 2013-02-23 &    cc-3 & $   8096 \pm 4   $ & $ 13.1 \pm 0.1 $ & $   10.4 \pm 0.1 $ &   B &      \\ 
                 & 2010-06-15 &    cc-4 & $   9258 \pm 12  $ & $205.1 \pm 0.1 $ & $    9.2 \pm 0.1 $ &   B &      \\ 
                 & 2013-02-23 &    cc-4 & $   9127 \pm 8   $ & $205.1 \pm 0.1 $ & $    9.5 \pm 0.1 $ &   B &      \\ 
                 & 2010-06-15 &    cc-5 & $   5434 \pm 10  $ & $ 22.6 \pm 0.1 $ & $   13.3 \pm 0.1 $ &   U &      \\ 
                 & 2010-06-15 &    cc-6 & $   6963 \pm 10  $ & $107.8 \pm 0.1 $ & $   13.9 \pm 0.2 $ &   U &      \\ 
                 & 2010-06-15 &    cc-7 & $   7158 \pm 12  $ & $130.1 \pm 0.1 $ & $   12.4 \pm 0.1 $ &   U &      \\ 
                 & 2010-06-15 &    cc-8 & $   7968 \pm 11  $ & $202.0 \pm 0.1 $ & $   13.9 \pm 0.3 $ &   U &      \\ 
                 & 2010-06-15 &    cc-9 & $   8756 \pm 9   $ & $ 12.5 \pm 0.1 $ & $   12.9 \pm 0.2 $ &   U &      \\ 
       HIP72399   & 2011-05-27   & none           & & & &    &  SB1, RV var\\
 TYC\,7835-2569-1 & 2011-05-11   & none           & & & &    & SB2 + Known binary$^b$ \\
      HIP76829   & 2010-06-15 &    cc-1 & $   2393 \pm 4   $ & $ 46.8 \pm 0.1 $ & $   15.0 \pm 0.4 $ &   B &      \\ 
                 & 2011-06-27 &    cc-1 & $   2705 \pm 9   $ & $ 45.7 \pm 0.2 $ & $   14.5 \pm 0.5 $ &   B &      \\ 
                 & 2010-06-15 &    cc-2 & $   6368 \pm 6   $ & $111.7 \pm 0.1 $ & $   14.8 \pm 0.3 $ &   B &      \\ 
                 & 2011-06-27 &    cc-2 & $   6434 \pm 10  $ & $109.1 \pm 0.1 $ & $   14.2 \pm 0.3 $ &   B &      \\
                 & 2010-06-15 &    cc-3 & $   4481 \pm 6   $ & $310.5 \pm 0.1 $ & $   16.3 \pm 0.6 $ &   U &      \\ 
                 & 2010-06-15 &    cc-4 & $   5272 \pm 3   $ & $272.4 \pm 0.1 $ & $   16.3 \pm 0.4 $ &   U &      \\ 
                 & 2010-06-15 &    cc-5 & $   5900 \pm 3   $ & $ 91.3 \pm 0.1 $ & $   16.2 \pm 0.4 $ &   U &      \\ 
                 & 2010-06-15 &    cc-6 & $   8618 \pm 11  $ & $142.8 \pm 0.1 $ & $   15.2 \pm 0.4 $ &   U &      \\ 
 TYC\,6781-0415-1 & 2011-07-20   & none           & & & &    &  \\
 TYC\,6786-0811-1 & 2010-07-29   & none           & & & &    &  Known binary$^c$\\
       HIP78747   & 2010-07-29 &    cc-1 & $   3892 \pm 2   $ & $  0.6 \pm 0.1 $ & $   11.9 \pm 0.1 $ &   B &      \\ 
                 & 2011-06-28 &    cc-1 & $   3990 \pm 4   $ & $  2.2 \pm 0.1 $ & $   12.0 \pm 0.1 $ &   B &      \\ 
                 & 2010-07-29 &    cc-2 & $   6154 \pm 5   $ & $292.2 \pm 0.1 $ & $   12.8 \pm 0.1 $ &   B &      \\ 
                 & 2011-06-28 &    cc-2 & $   6110 \pm 6   $ & $293.5 \pm 0.1 $ & $   13.0 \pm 0.2 $ &   B &      \\ 
                 & 2010-07-29 &    cc-3 & $   6633 \pm 8   $ & $ 53.9 \pm 0.1 $ & $   12.7 \pm 0.1 $ &   B &      \\ 
                 & 2011-06-28 &    cc-3 & $   6788 \pm 9   $ & $ 53.9 \pm 0.1 $ & $   12.8 \pm 0.2 $ &   B &      \\ 
                 & 2010-07-29 &    cc-4 & $   5508 \pm 6   $ & $ 59.0 \pm 0.1 $ & $   14.6 \pm 0.2 $ &   U &      \\ 
                 & 2010-07-29 &    cc-5 & $   5949 \pm 2   $ & $  1.1 \pm 0.1 $ & $   14.7 \pm 0.3 $ &   U &      \\ 
                 & 2010-07-29 &    cc-6 & $   7557 \pm 5   $ & $200.8 \pm 0.1 $ & $   14.0 \pm 0.2 $ &   U &      \\ 
                 & 2011-06-28 &    cc-7 & $   7419 \pm 5   $ & $255.3 \pm 0.1 $ & $   13.2 \pm 0.2 $ &   U &      \\ 

 TYC\,6209-0769-1 & 2011-08-19 &    cc-1 & $   5473 \pm 3   $ & $198.6 \pm 0.1 $ & $    8.2 \pm 0.1 $ &   U &      \\ 
       HIP79958   & 2011-06-27 &    cc-1 & $   3583 \pm 4   $ & $ 29.3 \pm 0.1 $ & $   11.7 \pm 0.3 $ &   U &      \\ 
                 & 2011-06-27 &    cc-2 & $   3689 \pm 1   $ & $ 88.8 \pm 0.1 $ & $    9.4 \pm 0.1 $ &   U &      \\ 
                 & 2011-06-27 &    cc-3 & $   4120 \pm 2   $ & $281.1 \pm 0.1 $ & $   10.2 \pm 0.1 $ &   U &      \\ 
                 & 2011-06-27 &    cc-4 & $   4633 \pm 4   $ & $151.3 \pm 0.1 $ & $   11.5 \pm 0.2 $ &   U &      \\ 
                 & 2011-06-27 &    cc-5 & $   5986 \pm 2   $ & $169.4 \pm 0.1 $ & $   10.9 \pm 0.1 $ &   U &      \\ 
                 & 2011-06-27 &    cc-6 & $   6195 \pm 1   $ & $177.4 \pm 0.1 $ & $   10.6 \pm 0.1 $ &   U &      \\ 
                 & 2011-06-27 &    cc-7 & $   7628 \pm 2   $ & $350.9 \pm 0.1 $ & $   11.3 \pm 0.3 $ &   U &      \\ 
       HIP80290  & 2011-08-08 &    cc-1 & $   2688 \pm 1   $ & $184.6 \pm 0.1 $ & $    8.5 \pm 0.1 $ &   B &      \\ 
                 & 2012-08-12 &    cc-1 & $   2665 \pm 7   $ & $184.3 \pm 0.2 $ & $    9.1 \pm 0.1 $ &   B &      \\ 
                 & 2011-08-08 &    cc-2 & $   3340 \pm 1   $ & $257.5 \pm 0.1 $ & $    1.9 \pm 0.1 $ &   C &      \\ 
                 & 2012-08-12 &    cc-2 & $   3335 \pm 8   $ & $257.6 \pm 0.2 $ & $    2.6 \pm 0.1 $ &   C & New binary (see Table~6)     \\ 
                 & 2011-08-08 &    cc-3 & $   7425 \pm 9   $ & $143.6 \pm 0.1 $ & $    6.9 \pm 0.1 $ &   B &      \\ 
                 & 2012-08-12 &    cc-3 & $   7417 \pm 12  $ & $143.4 \pm 0.1 $ & $    7.6 \pm 0.1 $ &   B &      \\ 
                 & 2012-08-12 &    cc-4 & $   2097 \pm 8   $ & $ 34.3 \pm 0.2 $ & $   12.4 \pm 0.3 $ &   U &      \\ 
                 & 2012-08-12 &    cc-5 & $   2245 \pm 8   $ & $291.3 \pm 0.2 $ & $   11.1 \pm 0.1 $ &   U &      \\ 
                 & 2012-08-12 &    cc-6 & $   6186 \pm 8   $ & $ 94.6 \pm 0.1 $ & $   12.0 \pm 0.1 $ &   U &      \\ 
                 & 2012-08-12 &    cc-7 & $   8629 \pm 11  $ & $298.8 \pm 0.1 $ & $   11.0 \pm 0.1 $ &   U &      \\ 
       HIP80758  & 2010-07-29 &    cc-1 & $   2210 \pm 3   $ & $163.6 \pm 0.1 $ & $   12.9 \pm 0.1 $ &   B &      \\ 
                 & 2011-05-11 &    cc-1 & $   2171 \pm 7   $ & $163.3 \pm 0.2 $ & $   12.2 \pm 0.2 $ &   B &      \\ 
                 & 2010-07-29 &    cc-2 & $   2221 \pm 4   $ & $241.1 \pm 0.1 $ & $   12.5 \pm 0.1 $ &   B &      \\ 
                 & 2011-05-11 &    cc-2 & $   2192 \pm 7   $ & $242.1 \pm 0.2 $ & $   11.8 \pm 0.2 $ &   B &      \\ 
                 & 2010-07-29 &    cc-3 & $   2413 \pm 4   $ & $321.1 \pm 0.1 $ & $   12.8 \pm 0.1 $ &   B &      \\ 
                 & 2011-05-11 &    cc-3 & $   2455 \pm 7   $ & $321.7 \pm 0.2 $ & $   12.3 \pm 0.2 $ &   B &      \\ 
                 & 2010-07-29 &    cc-4 & $   4686 \pm 6   $ & $236.1 \pm 0.1 $ & $   11.5 \pm 0.1 $ &   B &      \\ 
                 & 2011-05-11 &    cc-4 & $   4651 \pm 8   $ & $236.6 \pm 0.1 $ & $   10.9 \pm 0.1 $ &   B &      \\ 
                 & 2010-07-29 &    cc-5 & $   5228 \pm 7   $ & $304.1 \pm 0.1 $ & $   12.7 \pm 0.1 $ &   B &      \\ 
                 & 2011-05-11 &    cc-5 & $   5256 \pm 9   $ & $304.6 \pm 0.1 $ & $   12.1 \pm 0.1 $ &   B &      \\ 
                 & 2010-07-29 &    cc-6 & $   5229 \pm 5   $ & $ 70.3 \pm 0.1 $ & $   13.9 \pm 0.2 $ &   B &      \\ 
                 & 2011-05-11 &    cc-6 & $   5215 \pm 8   $ & $ 69.9 \pm 0.1 $ & $   12.9 \pm 0.2 $ &   B &      \\ 
 \noalign{\smallskip}\hline                  \noalign{\smallskip}
\end{tabular}
\end{center}
\begin{list}{}{}
\item[\scriptsize{(b):}] \scriptsize{Known binary (Brandner et al. 1996)}
\item[\scriptsize{(c):}] \scriptsize{Known binary (K\"ohler et al. 2000)}
\end{list}
\end{table*}

\begin{table*}[p]
\caption{Companion candidates characterization and identification (for multi-epoch
  observations). Table~10-cont.}
\begin{center}
\small
\begin{tabular}{llllllll}     
\noalign{\smallskip}
\noalign{\smallskip}\hline
Name-1           &     UT-Date  & Nb Cand.    &     Sep      &     PA    &  $\Delta H$   &    Status     &    Comments     \\
                 &              &                 & (mas)         & (deg)     & (mag)         &               &                 \\
\noalign{\smallskip}\hline  \noalign{\smallskip}
       HIP80758  & 2010-07-29 &    cc-7 & $   5441 \pm 5   $ & $108.0 \pm 0.1 $ & $   11.9 \pm 0.1 $ &   B &      \\ 
                 & 2011-05-11 &    cc-7 & $   5418 \pm 7   $ & $107.7 \pm 0.1 $ & $   10.9 \pm 0.1 $ &   B &      \\ 
                 & 2010-07-29 &    cc-8 & $   5489 \pm 6   $ & $ 25.2 \pm 0.1 $ & $   10.2 \pm 0.1 $ &   B &      \\ 
                 & 2011-05-11 &    cc-8 & $   5523 \pm 8   $ & $ 24.8 \pm 0.1 $ & $    9.6 \pm 0.1 $ &   B &      \\ 
                 & 2010-07-29 &    cc-9 & $   7495 \pm 4   $ & $ 79.1 \pm 0.1 $ & $    6.7 \pm 0.1 $ &   B &      \\ 
                 & 2011-05-11 &    cc-9 & $   7472 \pm 7   $ & $ 78.7 \pm 0.1 $ & $    5.7 \pm 0.1 $ &   B &      \\ 
                 & 2010-07-29 &   cc-10 & $   7925 \pm 3   $ & $265.4 \pm 0.1 $ & $   12.7 \pm 0.2 $ &   B &      \\ 
                 & 2011-05-11 &   cc-10 & $   7897 \pm 7   $ & $265.5 \pm 0.1 $ & $   12.4 \pm 0.3 $ &   B &      \\ 
                 & 2010-07-29 &   cc-11 & $   3005 \pm 4   $ & $ 22.0 \pm 0.1 $ & $   14.4 \pm 0.3 $ &   U &      \\ 
 TYC\,6818-1336-1& 2011-07-20 &    cc-1 & $   3382 \pm 10  $ & $302.7 \pm 0.2 $ & $    7.3 \pm 0.1 $ &   U &      \\ 
                 & 2011-07-20 &    cc-2 & $   5824 \pm 10  $ & $291.7 \pm 0.1 $ & $    5.5 \pm 0.1 $ &   U &      \\ 
                 & 2011-07-20 &    cc-3 & $   8914 \pm 14  $ & $ 52.1 \pm 0.1 $ & $    2.8 \pm 0.1 $ &   U &      \\ 
TYC\,6815-0084-1 & 2013-06-02 &   none  &                    &                  &                    &     & SB2? \\   
TYC\,6815-0874-1 & 2012-08-13 &    cc-1 & $   2094 \pm 16  $ & $229.6 \pm 0.4 $ & $   12.2 \pm 0.2 $ &   U &      \\ 
                 & 2012-08-13 &    cc-2 & $   2224 \pm 16  $ & $333.4 \pm 0.4 $ & $   11.9 \pm 0.2 $ &   U &      \\ 
                 & 2012-08-13 &    cc-3 & $   2713 \pm 16  $ & $280.7 \pm 0.4 $ & $   12.8 \pm 0.2 $ &   U &      \\ 
                 & 2012-08-13 &    cc-4 & $   2754 \pm 16  $ & $ 12.8 \pm 0.3 $ & $   11.1 \pm 0.1 $ &   U &      \\ 
                 & 2012-08-13 &    cc-5 & $   3801 \pm 16  $ & $171.0 \pm 0.3 $ & $    9.6 \pm 0.1 $ &   U &      \\ 
                 & 2012-08-13 &    cc-6 & $   4035 \pm 17  $ & $ 36.9 \pm 0.2 $ & $   12.1 \pm 0.1 $ &   U &      \\ 
                 & 2012-08-13 &    cc-7 & $   4940 \pm 17  $ & $164.2 \pm 0.2 $ & $   13.2 \pm 0.2 $ &   U &      \\ 
                 & 2012-08-13 &    cc-8 & $   6046 \pm 17  $ & $285.5 \pm 0.2 $ & $   12.9 \pm 0.2 $ &   U &      \\ 
                 & 2012-08-13 &    cc-9 & $   6569 \pm 17  $ & $ 22.6 \pm 0.2 $ & $   13.1 \pm 0.2 $ &   U &      \\ 
                 & 2012-08-13 &   cc-10 & $   8354 \pm 19  $ & $306.1 \pm 0.1 $ & $   12.4 \pm 0.2 $ &   U &      \\ 
                 & 2012-08-13 &   cc-11 & $   9053 \pm 19  $ & $125.0 \pm 0.1 $ & $   12.0 \pm 0.1 $ &   U &      \\ 
 TYC\,7362-0724-1 & 2010-06-16   &$57(+211)$       & & & &  & Electronic table \\
                  & 2011-05-11   &$57$        & & & & B(+U)  & Electronic table  \\
 TYC\,8728-2262-1 & 2011-08-25 &    cc-1 & $   2821 \pm 12  $ & $254.8 \pm 0.3 $ & $    9.1 \pm 0.1 $ &   U &      \\ 
                 & 2011-08-25 &    cc-2 & $   4449 \pm 13  $ & $ 27.5 \pm 0.2 $ & $   11.4 \pm 0.1 $ &   U &      \\ 
                 & 2011-08-25 &    cc-3 & $   6232 \pm 14  $ & $130.2 \pm 0.1 $ & $   12.1 \pm 0.2 $ &   U &      \\ 
                 & 2011-08-25 &    cc-4 & $   6399 \pm 12  $ & $ 99.3 \pm 0.2 $ & $    6.3 \pm 0.0 $ &   U &      \\ 
                 & 2011-08-25 &    cc-5 & $   6883 \pm 12  $ & $166.3 \pm 0.1 $ & $   12.4 \pm 0.1 $ &   U &      \\ 
       HIP86672   & 2010-06-16   &$261$        & & & &     & Electronic table\\
                  & 2011-08-25   & none        & & & &     &  \\
                  & 2013-04-25   & $80(+181)$         & & & & B(+U) & Electronic table \\ 
       HIP89829   & 2011-06-13   &$99$        & & & &    & Electronic table\\
                  & 2012-08-09   &$29(+70)$        & & & & B(+U)  &  Electronic table\\
       HIP93375   & 2010-06-14 &    cc-1 & $   3208 \pm 11  $ & $121.6 \pm 0.2 $ & $   13.8 \pm 0.2 $ &   B &      \\ 
                 & 2011-05-30 &    cc-1 & $   3175 \pm 9   $ & $120.5 \pm 0.2 $ & $   13.6 \pm 0.2 $ &   B &      \\ 
                 & 2010-06-14 &    cc-2 & $   4261 \pm 10  $ & $  9.2 \pm 0.2 $ & $   13.7 \pm 0.2 $ &   B &      \\ 
                 & 2011-05-30 &    cc-2 & $   4335 \pm 8   $ & $  9.0 \pm 0.2 $ & $   13.4 \pm 0.2 $ &   B &      \\ 
                 & 2010-06-14 &    cc-3 & $   4595 \pm 10  $ & $264.5 \pm 0.2 $ & $   12.5 \pm 0.1 $ &   B &      \\ 
                 & 2011-05-30 &    cc-3 & $   4591 \pm 8   $ & $265.7 \pm 0.2 $ & $   12.8 \pm 0.1 $ &   B &      \\ 
                 & 2010-06-14 &    cc-4 & $   4822 \pm 12  $ & $139.6 \pm 0.1 $ & $   11.4 \pm 0.1 $ &   B &      \\ 
                 & 2011-05-30 &    cc-4 & $   4754 \pm 10  $ & $138.9 \pm 0.1 $ & $   11.5 \pm 0.1 $ &   B &      \\ 
                 & 2010-06-14 &    cc-5 & $   5308 \pm 10  $ & $173.7 \pm 0.2 $ & $   13.7 \pm 0.2 $ &   B &      \\ 
                 & 2011-05-30 &    cc-5 & $   5218 \pm 8   $ & $173.6 \pm 0.1 $ & $   13.5 \pm 0.1 $ &   B &      \\ 
                 & 2010-06-14 &    cc-6 & $   5354 \pm 10  $ & $176.9 \pm 0.2 $ & $   14.0 \pm 0.2 $ &   B &      \\ 
                 & 2011-05-30 &    cc-6 & $   5283 \pm 8   $ & $177.1 \pm 0.1 $ & $   14.0 \pm 0.2 $ &   B &      \\ 
                 & 2010-06-14 &    cc-7 & $   6095 \pm 10  $ & $274.5 \pm 0.1 $ & $   12.0 \pm 0.1 $ &   B &      \\ 
                 & 2011-05-30 &    cc-7 & $   6104 \pm 8   $ & $275.2 \pm 0.1 $ & $   12.7 \pm 0.1 $ &   B &      \\ 
                 & 2010-06-14 &    cc-8 & $   6848 \pm 10  $ & $185.1 \pm 0.1 $ & $   12.4 \pm 0.1 $ &   B &      \\ 
                 & 2011-05-30 &    cc-8 & $   6763 \pm 8   $ & $184.9 \pm 0.1 $ & $   11.6 \pm 0.1 $ &   B &      \\ 
                 & 2010-06-14 &    cc-9 & $   6942 \pm 12  $ & $153.9 \pm 0.1 $ & $   12.4 \pm 0.1 $ &   B &      \\ 
                 & 2011-05-30 &    cc-9 & $   6856 \pm 10  $ & $154.2 \pm 0.1 $ & $   11.6 \pm 0.1 $ &   B &      \\ 
                 & 2010-06-14 &   cc-10 & $   7089 \pm 14  $ & $ 46.6 \pm 0.1 $ & $   12.9 \pm 0.1 $ &   B &      \\ 
                 & 2011-05-30 &   cc-10 & $   7143 \pm 12  $ & $ 45.8 \pm 0.1 $ & $   13.5 \pm 0.2 $ &   B &      \\ 
                 & 2010-06-14 &   cc-11 & $   7502 \pm 13  $ & $116.6 \pm 0.1 $ & $   11.1 \pm 0.1 $ &   U &      \\ 
                 & 2010-06-14 &   cc-12 & $   7512 \pm 12  $ & $115.1 \pm 0.1 $ & $   12.6 \pm 0.1 $ &   U &      \\ 
                 & 2011-05-30 &   cc-13 & $   4157 \pm 9   $ & $237.0 \pm 0.1 $ & $   14.1 \pm 0.2 $ &   U &      \\ 
                 & 2011-05-30 &   cc-14 & $   5917 \pm 8   $ & $262.2 \pm 0.1 $ & $   14.4 \pm 0.3 $ &   U &      \\ 
                 & 2011-05-30 &   cc-15 & $   9234 \pm 14  $ & $322.3 \pm 0.1 $ & $   11.5 \pm 0.1 $ &   U &      \\ 
  \noalign{\smallskip}\hline                  \noalign{\smallskip}
\end{tabular}
\end{center}
\end{table*}

\begin{table*}[p]
\caption{Companion candidates characterization and identification (for multi-epoch
  observations). Table~11-cont.}
\begin{center}
\small
\begin{tabular}{llllllll}     
\noalign{\smallskip}
\noalign{\smallskip}\hline
Name-1           &     UT-Date  & Nb Cand.    &     Sep      &     PA    &  $\Delta H$   &    Status     &    Comments     \\
                 &              &                 & (mas)         & (deg)     & (mag)         &               &                 \\
\noalign{\smallskip}\hline  \noalign{\smallskip}
      HIP94235   & 2010-07-30   &        none  & & & &    &  New binary (see Table~6) \\
 TYC\,6893-1391-1 & 2011-06-08 &    cc-1 & $   3289 \pm 4   $ & $229.6 \pm 0.1 $ & $   11.2 \pm 0.2 $ &   U &      \\ 
                 & 2011-06-08 &    cc-2 & $   3373 \pm 1   $ & $256.6 \pm 0.1 $ & $   11.8 \pm 0.3 $ &   U &      \\ 
                 & 2011-06-08 &    cc-3 & $   5761 \pm 7   $ & $224.5 \pm 0.1 $ & $    6.8 \pm 0.0 $ &   U &      \\ 
 TYC\,5206-0915-1 & 2010-07-30   &        none  & & & &    &  \\
TYC\,5736-0649-1  & 2011-08-18 &    cc-1 & $   4360 \pm 6   $ & $206.3 \pm 0.1 $ & $    9.8 \pm 0.1 $ &   U &      \\ 
                  & 2011-08-18 &    cc-2 & $   6130 \pm 8   $ & $306.9 \pm 0.1 $ & $   10.6 \pm 0.1 $ &   U &      \\ 
       HD189285   & 2011-08-20 &    cc-1 & $   4519 \pm 4   $ & $ 24.9 \pm 0.1 $ & $    9.5 \pm 0.1 $ &   U &      \\ 
       HIP98470   & 2010-06-15 &        none  & & & &    &  \\
TYC\,5164-567-1   & 2011-07-29 &    cc-1 & $   2632 \pm 3   $ & $207.5 \pm 0.1 $ & $    3.3 \pm 0.0 $ &   U &      \\ 
                 & 2011-07-29 &    cc-2 & $   4421 \pm 5   $ & $ 56.8 \pm 0.1 $ & $   11.2 \pm 0.1 $ &   U &      \\ 
                 & 2011-07-29 &    cc-3 & $   5674 \pm 7   $ & $229.3 \pm 0.1 $ & $    8.6 \pm 0.1 $ &   U &      \\ 
                 & 2011-07-29 &    cc-4 & $   7254 \pm 9   $ & $139.8 \pm 0.1 $ & $    9.8 \pm 0.1 $ &   U &      \\ 
       HIP99273   & 2010-07-31   &        none  & & & &    &  \\
       HD199058   & 2010-06-15   &        none  & & & &    &  New binary (see Table~6) \\
      HIP105384   & 2010-07-31   & none           & & & &    &  \\
                  & 2011-06-08 &    cc-1 & $   7038 \pm 7   $ & $ 24.5 \pm 0.1 $ & $   14.1 \pm 0.2 $ &   U &      \\ 
      HIP105612   & 2010-07-31   &        none  & & & &    &  \\
      HIP107684   & 2010-06-15   &        none  & & & &    & New binary (see Table~6) \\
      HIP108422   & 2010-07-30   &        none  & & & &    & Known binary$^d$  \\
 TYC\,8004-0083-1 & 2010-06-15 &        none & & &    &  \\
      HIP114046   & 2010-06-15 &        none & & &    &  \\
 TYC\,9338-2016-1 & 2009-11-23 &        none & & &    &  \\
                  & 2010-07-30 &        none & & &    &  \\
 TYC\,9529-0340-1 & 2010-07-31 &        none & & &    &  \\
 TYC\,9339-2158-1 & 2010-07-31 &        none & & &    &  \\
 TYC\,6406-0180-1 & 2010-07-30 &        none & & &    &  \\
      HIP116910   & 2009-11-22 &        none & & &    &  \\
  \noalign{\smallskip}\hline                  \noalign{\smallskip}
\end{tabular}
\begin{list}{}{}
\item[\scriptsize{(d):}] \scriptsize{Known binary (Chauvin et al. 2010)}
\end{list}
\end{center}
\end{table*}


\end{document}